\documentclass[iop,apj,numberedappendix,appendixfloats]{emulateapj}

\usepackage{amsmath}
\usepackage{multirow}
\usepackage{graphicx,float}

\bibpunct[,]{(}{)}{;}{a}{}{,}

\newcommand{\beq}{\begin{equation}}
  \newcommand{\eeq}{\end{equation}}
\newcommand{\be}{\begin{equation}}
  \newcommand{\ee}{\end{equation}}

\newcommand{\hgvol}{h^{-3}{\mathrm{Gpc}}^{3}}
\newcommand{\hgpc}{h^{-1}\mathrm{Gpc}}
\newcommand{\hmpc}{h^{-1}\mathrm{Mpc}}
\newcommand{\hkpc}{h^{-1}\mathrm{kpc}}
\newcommand{\hMsun}{h^{-1}M_{\odot}}

\newcommand{\fnl}{f_\mathrm{NL}}

\newcommand{\hden}{\; h^{3}{\mathrm{Mpc}}^{-3}}

\newcommand{\nbody}{$N$-body}

\begin{document}

\title{Constraining Primordial Non-Gaussianity with Moments of the Large Scale Density Field}

\author{
  Qingqing~Mao\altaffilmark{1},
  Andreas~A.~Berlind\altaffilmark{1},
  Cameron~K.~McBride\altaffilmark{2},
  Robert~J.~Scherrer\altaffilmark{1},
  Rom\'an Scoccimarro\altaffilmark{3},
  Marc Manera\altaffilmark{4},
}
\altaffiltext{1}{Department of Physics and Astronomy, Vanderbilt University, Nashville, TN 37235, USA}
\altaffiltext{2}{Harvard-Smithsonian Center for Astrophysics, Cambridge, MA 02138, USA}
\altaffiltext{3}{Center for Cosmology and Particle Physics \& Department of Physics, New York University, New York, NY 10003, USA}
\altaffiltext{4}{University College London, Gower Street, London WC1E 6BT, UK}

\begin{abstract}
We use cosmological N-body simulations to investigate whether measurements of the moments of large-scale structure can yield constraints on primordial non-Gaussianity. We measure the variance, skewness, and kurtosis of the evolved density field from simulations with Gaussian and three different non-Gaussian initial conditions: a local model with $\fnl=100$, an equilateral model with $\fnl=-400$, and an orthogonal model with $\fnl=-400$. We show that the moments of the dark matter density field differ significantly between Gaussian and non-Gaussian models. We also make the measurements on mock galaxy catalogs that contain galaxies with clustering properties similar to those of luminous red galaxies (LRGs). We find that, in the case of skewness and kurtosis, galaxy bias reduces the detectability of non-Gaussianity, though we can still clearly discriminate between different models in our simulation volume. However, in the case of the variance, galaxy bias greatly amplifies the detectability of non-Gaussianity. In all cases we find that redshift distortions do not significantly affect the detectability.  When we restrict our measurements to volumes equivalent to the Sloan Digital Sky Survey II (SDSS-II) or Baryon Oscillation Spectroscopic Survey (BOSS) samples, the probability of detecting a departure from the Gaussian model is high by using measurements of the variance, but very low by using only skewness and kurtosis measurements. For example, if our local non-Gaussian model were the true model in the universe, a variance measurement in the BOSS survey would have a $\sim$95\% chance of detecting this non-Gaussiantity at the $2\sigma$ level, whereas a skewness measurement would only have at best a $\sim$25\% chance of doing so. We estimate that in order to detect an amount of non-Gaussianity that is consistent with recent CMB constraints using skewness or kurtosis, we would need a galaxy survey that is much larger than any planned future survey. Skewness and kurtosis measurements are thus never likely to yield useful constraints on primordial non-Gaussianity. On the other hand, future surveys should be large enough to place meaningful constraints using measurements of the galaxy variance.
\end{abstract}


\section{Introduction}
\label{s:introduction}

Inflation is the most promising paradigm for the early universe \citep{Guth:1981}. The standard inflationary paradigm predicts nearly Gaussian and scale invariant primordial density fluctuations, which are consistent with the observations of the Cosmic Microwave Background (CMB) and Large-Scale Structure (LSS) in the last few decades. However, even the simplest inflation model predicts some small deviation from Gaussianity \citep{Falk:1993, Gangui:1994, Maldacena:2003, Bartolo:2004}. Within the standard inflationary paradigm, there are currently many viable inflationary models, but it is difficult to discriminate between them. While most of the popular inflation models predict slight deviations from Gaussian fluctuations, different models predict different amounts and flavors of non-Gaussianity, which makes it a very powerful tool for constraining inflationary models (e.g., see \citealt{Chen:2010} for a review). Detecting primordial non-Gaussianity is thus an important goal of modern cosmology and it has recently garnered much attention.

The primordial density fluctuations are both the direct cause of CMB anisotropy, and the seeds of large scale structure (LSS) formation. Deviations from primordial Gaussianity can thus leave signals on both the CMB and LSS. To date, observations of the CMB have been playing the central role in constraining the amplitudes of various types of primordial non-Gaussianity, with tight constraints coming from both WMAP \citep{Bennett:2013} and, most recently, Planck \citep[][ paper XXIV]{PlanckCollaboration:2013}. However, ongoing and future high quality redshift surveys raise hope for detecting non-Gaussianity in LSS. The Sloan Digital Sky Survey (SDSS; \citealt{York:2000}) has provided redshifts of over 100,000 luminous red galaxies (LRGs) in a large volume \citep{Eisenstein:2001}, and the ongoing Baryon Oscillation Spectroscopic Survey (BOSS; \citealt{Dawson:2013}), which is part of the SDSS-III project \citep{Eisenstein:2011}, is mapping 1.5 million luminous galaxies to redshift $z \sim 0.7$. Future redshift surveys like eBOSS, DESI, and Euclid will map even larger volumes.  These surveys provide great opportunities of constraining primordial non-Gaussianity with large-scale structure.

There are several avenues for constraining primordial non-Gaussianity with galaxy surveys, including the galaxy power spectrum, higher order correlations of the density field, e.g., the bispectrum, and statistics of rare peaks, i.e., the abundance of massive clusters. There have been many studies attempting to detect non-Gaussianity using the galaxy power spectrum \cite[e.g., ][]{Slosar:2008,Afshordi:2008,Ross:2013,Giannantonio:2013}. The galaxy bispectrum is much more difficult to measure and there has only been one attempt to use it for the purpose of constraining non-gaussianity \citep{Scoccimarro:2004}. However, it provides a highly sensitive probe of non-Gaussianity and is likely to yield the best constraints from LSS with future surveys \citep{Sefusatti:2007, Sefusatti:2009, Baldauf:2011, Scoccimarro:2012}

A much simpler set of statistics for quantifying departures from Gaussianity are the higher order moments of the density field, of which the most frequently used are the third order normalized moment \emph{skewness} and fourth order normalized moment \emph{kurtosis}. Though gravitational evolution contributes most of the signal in these moments in the present day density field, small departures from Gaussianity in the primordial density field may still cause slightly different skewness and kurtosis today, which may be detectable in sufficiently large galaxy redshift surveys.

The evolution of skewness, and kurtosis for Gaussian initial conditions has been studied both analytically and numerically in many published works \citep{Peebles:1980, Fry:1985, Coles:1991, Juszkiewicz:1992, Weinberg:1992, Bouchet:1992, Lahav:1993, Luo:1993, Coles:1993, Juszkiewicz:1993, Lucchin:1994, Frieman:1994, Bernardeau:1994, Hui:1999, Bernardeau:2002}. For arbitrary non-Gaussian initial conditions, \cite{Fry:1994} computed the evolution of skewness in second-order perturbation theory, and \cite{Chodorowski:1996} computed the kurtosis case. Observationally, skewness and kurtosis have been measured for many galaxy redshift surveys \citep{Bernardeau:2002}, such as QDOT \citep{Saunders:1991}, 1.2Jy IRAS \citep{Bouchet:1992, Bouchet:1993, Kim:1998}, CfA-SRSS \citep{Gaztanaga:1992}, 1.9Jy IRAS \citep{Fry:1994a}, PPS \citep{Ghigna:1996}, SRSS2 \citep{Benoist:1999}, PSCz \citep{Szapudi:2000p}, Durham/UKST \citep{Hoyle:2000}, Stromlo/APM \citep{Hoyle:2000}, 2dFGRS \citep{Croton:2004}, VVDS \citep{Marinoni:2005}, and SDSS \citep{Szapudi:2002, Ross:2008, P'apai:2010}. So far all results are consistent with Gaussian initial conditions, but these surveys have not had sufficient volume to detect plausible amounts of primordial non-Gaussianity.

With much larger redshift surveys coming out in the next decade, we think this is a good time to revisit this question. Though the skewness and kurtosis contain less information than their corresponding non-zero separation correlations, the 3 and 4-point correlation functions (and their Fourier transforms, the bispectrum and trispectrum), they are conceptually simpler and much easier to measure. In this paper, we use N-body simulations to investigate the detectability of inflationary-motivated primordial non-Gaussianity from skewness and kurtosis measurements of the present day galaxy distribution. We also investigate the second order moment of the density field, \emph{variance}, which contains similar information to the power spectrum. In \S\ref{s:theory}, we review the background theory and some related definitions. In \S\ref{s:data} we present the details of our simulations, which include both Gaussian and non-Gaussian initial conditions, and we describe how we measure the density field moments from these simulations. We show our results in \S\ref{s:results}, including measurements on both dark matter particles and mock galaxy catalogs constructed to model the distribution of SDSS LRGs. We also make measurements on subsets of the simulations that have volumes equivalent to the SDSS-II and BOSS surveys, and we calculate the likelihood of detecting departures from the Gaussian model with variance, skewness or kurtosis measurements from these surveys. We present our conclusions and some discussion in \S\ref{s:discussion}.

\section{Background Theory}
\label{s:theory}

\subsection{Skewness and Kurtosis}

The smoothed density fluctuation $\delta_R$ with smoothing scale $R$ can be written as
\begin{equation}
  \delta_R=\frac{\rho_R}{\langle \rho_R \rangle}-1 ,
\end{equation}
where $\rho_R$ is the smoothed density. The \emph{variance} of the density field is $\langle \delta_R^2\rangle$. Higher order moments are typically normalized by the variance so that the normalized moment of order $n$ is defined as
\begin{equation}
  s_n \equiv \frac{\langle \delta_R^n \rangle_c}{\langle \delta_R^2 \rangle^{n/2}} ,
\end{equation}
while $\langle \delta_R^n \rangle_c$ is the $n$th order connected moment. The third and fourth order normalized moments are called \emph{skewness} and \emph{kurtosis} respectively. Another definition commonly used is the \emph{hierarchical amplitude}:
\begin{equation}
  S_n \equiv \frac{\langle \delta_R^n \rangle_c}{\langle \delta_R^2 \rangle_c^{n-1}} .
  \label{Eq:Sn}
\end{equation}
In the literature of large-scale structure, the third and fourth hierarchical amplitudes $S_3$ and $S_4$ are often referred to as the skewness and kurtosis parameters, respectively. Hereafter in this paper, we also use this definition:
\begin{equation}
  S_3 = \frac{\langle \delta_R^3 \rangle_c}{\langle \delta_R^2 \rangle_c^2} = \frac{\langle \delta_R^3 \rangle}{\langle \delta_R^2 \rangle^2} ,
  \label{Eq:S3}
\end{equation}
\begin{equation}
  S_4 = \frac{\langle \delta_R^4 \rangle_c}{\langle \delta_R^2 \rangle_c^3} = \frac{\langle \delta_R^4 \rangle - 3\langle \delta_R^2 \rangle^2}{\langle \delta_R^2 \rangle^3}
  \label{Eq:S4} .
\end{equation}

For a Gaussian initial distribution, second-order perturbation theory predicts constant values for these parameters, with $S_3 = 34/7 $ \citep{Peebles:1980} and $S_4 = 60712/1323 $ \citep{Bernardeau:1992}, if smoothing is not considered. Including the effect of top-hat smoothing, $S_3$ and $S_4$ can also be derived and have the form \citep{Bernardeau:1994}
\begin{equation}
  S_3=\frac{34}{7}+\gamma_1 ,
  \label{Eq:S3theory}
\end{equation}
\begin{equation}
  S_4=\frac{60712}{1323}+\frac{62\gamma_1}{3}+\frac{7\gamma_1^2}{3}+\frac{2\gamma_2}{3} ,
  \label{Eq:S4theory}
\end{equation}
where
\begin{equation}
  \gamma_p=\frac{d^p \log{\sigma^2(R)}}{d \log^p R} ,
\label{Eq:gamma}
\end{equation}
and $\sigma^2(R)$ is another way of denoting the variance of the density field smoothed on a scale $R$.

\subsection{Non-Gaussian Initial Distribution}

To describe primordial non-Gaussianity generated during inflation, the initial conditions are commonly written as the sum of a linear Gaussian term and a non-linear quadratic term that contains the deviation from Gaussianity:
\begin{equation}
  \Phi=\phi+\frac{\fnl}{c^2}(\phi^2-\langle\phi^2\rangle).
  \label{Eq:fnl}
\end{equation}
Here $\Phi$ is Bardeen's gauge-invariant potential \citep{Salopek:1990}, and $\phi$ denotes a Gaussian random field. In general, the dimensional parameter $\fnl$ is scale and configuration dependent. When $\fnl$ is simply a constant, it yields the so called \emph{local} model. In Fourier space, the bispectrum of the local model can be written as \citep{Gangui:1994,Verde:2000,Komatsu:2001}
\begin{equation}
  B_{local}(k_1,k_2,k_3)= 2 \fnl [ P(k_1)P(k_2) + 2cyc. ] .
  \label{Eq:local}
\end{equation}
Here $P(k)$ is the power spectrum and $cyc.$ denotes the cyclic terms over $k_1,k_2,k_3$. The bispectrum for the local type non-Gaussianity peaks when $k_1 \approxeq k_2 \ll k_3$ (the so-called ``squeezed'' configuration; \citealt{Babich:2004}).

While beyond single-field models of inflation generically predict the local type, single-field models generate predominantly other forms. One is the \emph{equilateral} type, for which the bispectrum peaks when $k_1 \simeq k_2 \simeq k_3$, and can be written as \citep{Creminelli:2006}
\begin{equation}
  \begin{aligned}
    B_{equil}(k_1,k_2,k_3) = 6 \fnl [ -P(k_1)P(k_2) + 2cyc.\\
    - 2[P(k_1)P(k_2)P(k_3)]^{2/3}\\
    + P^{1/3}(k_1)P^{2/3}(k_2)P(k_3) + 5cyc.] .
    \label{Eq:equilateral}
  \end{aligned}
\end{equation}
\cite{Senatore:2010} constructed another distinct shape of non-Gaussianity called \emph{orthogonal}, for which the bispectrum can be approximately given by this template:
\begin{equation}
  \begin{aligned}
    B_{orthog}(k_1,k_2,k_3) = 6 \fnl [ -3P(k_1)P(k_2) + 2cyc.\\
    - 8[P(k_1)P(k_2)P(k_3)]^{2/3}\\
    + 3P^{1/3}(k_1)P^{2/3}(k_2)P(k_3) + 5cyc.] .
    \label{Eq:orthogonal}
  \end{aligned}
\end{equation}
More precisely, this template is only a good approximation to the orthogonal shape away from the squeezed limit ($k_3 \to 0$). This is relevant to the calculation of the large-scale bias, as the more accurate template does not lead to a scale-dependent bias at low-$k$ whereas the simpler template in equation~(\ref{Eq:orthogonal}) leads to a $1/k$ correction to the  bias. On the other hand, such a behavior is interesting from a phenomenological point of view as it is in between scale independence and the $1/k^2$ of local PNG.
In this paper, we use N-body simulations that are generated with all three of the above types of non-Gaussian initial conditions, and are described in detail by \cite{Scoccimarro:2012}.

The evolution of skewness for non-Gaussian initial conditions was first investigated by \cite{Fry:1994}; this was extended to the kurtosis by \cite{Chodorowski:1996}.  Although general expressions for $S_3$ and $S_4$ can be derived for arbitrary non-Gaussian initial conditions, these generally involve complicated integrals over the initial density field correlators and are not easily generalized to the smoothed density field.  For $S_3$, for example, the general expression consists of a term encoding the initial (non-Gaussian) value for $S_3$, which decays as $1/a$, where $a$ is the scale factor, a second ``Gaussian" term which is constant and equal to the Gaussian value for $S_3$, and a third set of terms that are also constant and depend on the initial 3-point and 4-point correlations in the initial density field.

For the local non-Gaussian model, \citet{Scoccimarro:2004} derived an expression for the evolved bispectrum.  Similar to the derivation in \citet{Fry:1994}, the evolved bispectrum contains a ``Gaussian" piece identical to the bispectrum for Gaussian initial conditions, a ``non-Gaussian" piece corresponding to the non-Gaussian initial value of $S_3$, and a third piece arising from the trispectrum. \citet{Scoccimarro:2004} noted that the second term scales as $\fnl$, while the third scales as $\fnl^2$.  Thus, in the limit of small $\fnl$, it is sufficient to consider only the contributions from the Gaussian term and the term arising from the initial skewness.  These terms can be integrated with the appropriate window functions to give a reasonable estimate for $S_3$ for non-Gaussian initial conditions \citep{Scoccimarro:2004, Lam:2009a, Lam:2009b}. The value of $S_3$ in the evolved density field then involves a competition between the intrinsic initial value for $S_3$, which dominates on large scales, and the evolved Gaussian piece, which dominates on small scales.  Of course, in either case the expressions derived from quasi-linear perturbation theory become progressively less accurate on smaller (more nonlinear) scales.  In this paper, we only compare our results to the analytic expressions for $S_3$ and $S_4$ in the case of Gaussian initial conditions, i.e., equations (\ref{Eq:S3theory}) and (\ref{Eq:S4theory}). We only mention the non-Gaussian analytic expressions for the insight that they offer into our numerical results.

\subsection{Galaxy Bias}

In redshift surveys we observe galaxies, and their distribution is ``biased'' relative to the underlying mass distribution. We assume that the smoothed galaxy density fluctuation is a local function of the smoothed mass density fluctuation and can be expressed as a Taylor series:
\beq
\delta_{R}^{gal} = \sum\limits_{k=0}^{\infty} \frac{b_k}{k!}\delta_R^k .
\eeq
Here $b_0$ is fixed to be $b_0 = -\sum_{k=2}^{\infty} b_k\langle\delta_R^k\rangle/k!$ to make sure $\langle\delta_{R}^{gal}\rangle = 0$. The $b_1$ term corresponds to the usual linear bias factor $b$, and $b_2$ is the nonlinear quadratic bias. These bias factors are scale dependent on small scales, but they become scale independent on large scales \citep{Manera:2011}. By using the hierarchical relation equation (\ref{Eq:Sn}), the relation between the skewness and kurtosis parameters for galaxies and mass can be derived as \citep{Fry:1993}:
\beq
\label{Eq:biasS3}
S_{3}^{gal} = b^{-1}(S_3 + 3c_2),
\eeq
\beq
S_{4}^{gal} = b^{-2}(S_4 + 12c_2S_3 + 4c_3 + 12c_2^2) ,
\eeq
where $c_k = b_k/b$ for $k\geq2$.

Now let us consider the effect of biasing on the measurement of the variance and $S_3$ for non-Gaussian initial conditions. The total linear bias can be written as the usual (Gaussian) bias $b_G$ plus a non-Gaussian correction:
\beq
b_{NG} = b_G + \Delta b_{\fnl}.
\eeq
In general, the non-Gaussian correction of bias depends not only on $\fnl$, but also on scale. \citet{Dalal:2008} showed that for the local non-Gaussian model the linear bias correction depends on scale $k$ as
\beq
\label{Eq:biascorrection}
\Delta b_{\fnl}(k) = 2 ( b_G - 1 ) \fnl \delta_c \frac{3 \Omega_m}{2 a g(a) r^2_H k^2},
\eeq
where $\Omega_m$ is the matter density parameter, $a$ is the scale factor, $r_H$ is the Hubble radius, $\delta_c$ is the critical threshold for collapse, and $g(a)$ is the growth suppression rate defined as $D(a)/a$, where $D(a)$ is the growth factor. A similar scale dependence due to $f_{\rm NL}$ holds for the quadratic bias factor $b_2$ \citep{Giannantonio:2010, Scoccimarro:2012}, which must be considered when discussing $S_3$. These results give corrections to the bias in Fourier space that are scale ($k$) dependent and they can be used to calculate the bispectrum for the local non-Gaussian case. One can then integrate the resulting bispectrum numerically to obtain the skewness at a scale $R$.

\subsection{Discrete Distribution}
\label{ss:discrete}

Whether we measure density using dark matter particles or galaxies, in a simulation we always deal with discrete numbers of points. Specifically, we measure density by counting the number of dark matter particles or galaxies within top-hat spheres. The density fluctuation with smoothing radius $R$ is then
\beq
\delta_R = \frac{N}{\langle N \rangle} - 1 ,
\eeq
where $N$ is the number of particles in a given sphere and $\langle N \rangle$ is the mean over all spheres. Since counts are discrete numbers, we cannot directly use equations (\ref{Eq:S3}) and (\ref{Eq:S4}) to calculate $S_3$ and $S_4$. Here we apply a Poisson correction \citep{Peebles:1980} using the \cite{Lahav:1993} notation. The moments of the density fluctuation $\delta_R$ can be expressed in terms of the n-point correlation functions and Poisson terms involving $\langle N \rangle$:
\beq
\label{Eq:d2p}
\langle \delta_R^2 \rangle = \frac{1}{\langle N \rangle} + \Psi_2 ,
\eeq
\beq
\label{Eq:d3p}
\langle \delta_R^3 \rangle = \frac{1}{\langle N \rangle^2} + \frac{3}{\langle N \rangle}\Psi_2 + \Psi_3 ,
\eeq
\beq
\label{Eq:d4p}
\langle \delta_R^4 \rangle = \frac{1}{\langle N \rangle^3} + \frac{1}{\langle N \rangle^2}(3+7\Psi_2) + \frac{6}{\langle N \rangle}(\Psi_2+\Psi_3) + 3\Psi_2^2 + \Psi_4 ,
\eeq
where
\beq
\Psi_2 = \frac{1}{V^2} \int \xi_{12} dV_1 dV_2 ,
\eeq
\beq
\Psi_3 = \frac{1}{V^3} \int \zeta_{123} dV_1 dV_2 dV_3 ,
\eeq
\beq
\Psi_4 = \frac{1}{V^4} \int \eta_{1234} dV_1 dV_2 dV_3 dV_4 .
\eeq
$\xi_{12}$, $\zeta_{123}$, $\eta_{1234}$ denote two-, three- and four-point correlation functions, and $V$ is the volume of the smoothing sphere.

We can measure the moments of $\delta_R$ and the mean counts $\langle N \rangle$ directly from the simulations, solve for $\Psi_2$, $\Psi_3$ and $\Psi_4$ using equations~\ref{Eq:d2p}--\ref{Eq:d4p}, and then evaluate $S_3$ and $S_4$ as
\beq
\label{Eq:S3p}
S_3 = \frac{\Psi_3}{\Psi_2^2},
\eeq
\beq
\label{Eq:S4p}
S_4 = \frac{\Psi_4}{\Psi_2^3}.
\eeq
These are the main equations we use to measure the variance, skewness, and kurtosis parameters in this paper.

\section{Simulated Data}
\label{s:data}

\subsection{LasDamas Simulations}
\label{s:lasdamas}

We use simulated data from the Large Suite of Dark Matter Simulations project \citep[LasDamas;][]{McBride:2009}.  The LasDamas project has focused on running many independent \nbody\ realizations with the same cosmology but different initial phases. The simulation data we analyze have WMAP5 motivated cosmological parameters, specifically $\Omega_m=0.25, \Omega_\Lambda=0.75, \Omega_b=0.04, h=0.7, \sigma_8=0.8, n_s=1.0$.  The LasDamas simulations are designed to model SDSS galaxies and contain four different volume and resolution configurations that were chosen to match different luminosity samples. In this paper we focus on the the largest volume ``Oriana'' realizations, which are designed to model SDSS LRGs.  Each Oriana simulation evolves $1280^3$ dark matter particles in a cubic volume of $2.4 \hgpc$ on a side, resulting in a particle mass of $45.7 \times 10^{10} \hMsun$. The simulations are seeded with second-order Lagrangian perturbation theory (2LPT) initial conditions \citep{Roman:1998,Crocce:2006} and evolved from a starting redshift of $z_{init} = 49$ to $z=0$ using the Gadget-2 code \citep{Springel:2005}, with a gravitational force softening of $53 \hkpc$.  

The initial density field for the LasDamas simulations is Gaussian, and here we analyze $40$ Oriana realizations (over $550 \hgvol$ total volume).  To complement these simulations, we also have sets of simulations seeded with three different models of primordial non-Gaussianity: local, equilateral and orthogonal. These simulations are described in detail by \cite{Scoccimarro:2012}.  Specifically, we have 12 realizations of each non-Gaussian model, and these are constrained to have the same box size, resolution and initial phases as 12 of the Gaussian Oriana realizations.  We can thus compare the Gaussian and three non-Gaussian models in 12 boxes ($165 \hgvol$ total volume per model) without having to worry about cosmic variance differences.

Our non-Gaussian models have $\fnl$ amplitudes of 100 for the local model, and $-400$ for each of the equilateral and orthogonal models. These values were marginally consistent with constraints from WMAP at the time that the simulations were run. However, the recent Planck constraints have ruled these models out definitively, since they constrain $\fnl$ to be consistent with zero with $1\sigma$ errors of 5.8, 75, and 39 for the local, equilateral, and orthogonal models, respectively \citep{PlanckCollaboration:2013}. It is thus important to emphasize that the results that we present in this paper apply to our specific models and are exaggerated with respect to realistic models. In \S~\ref{ss:scaling_fnl}, we discuss how some of our conclusions might scale to much lower amplitude non-Gaussian models that are still allowed by the Planck constraints.

\subsection{Mock Galaxy Catalogs}
\label{ss:mocks}

To include the effects of galaxy bias, we analyze mock galaxy catalogs that model SDSS-II LRG galaxies. Specifically, we use two sets of mock catalogs from LasDamas that correspond to LRG samples with $g$-band absolute magnitudes of $M_g<-21.2$ and LRG $M_g<-21.8$.  The average comoving number density of these samples is $9.7 \times 10^{-5} \hden$ and $2.4 \times 10^{-5} \hden$, respectively \citep{Zehavi:2005, Kazin:2010}.  The mock catalogs were constructed by first identifying friends-of-friends halos in the dark matter distribution at $z=0.34$ (roughly the median redshift of the brighter LRG sample), and then populating these halos with galaxies using a halo occupation distribution (HOD; \citealt{Berlind:2002}).  The parameters of the HOD were determined by fitting to the observed small-scale clustering of LRGs, as described by \citet{McBride:2009}.  In each halo, a central galaxy was placed at the halo center and given the halo's mean velocity, and satellite galaxies were given the positions and velocities of randomly selected dark matter particles within the halo. We do not apply realistic observational sky footprints to our mock catalogs within the analysis we present here, but rather use the whole simulation cubes. To include the effects of redshift space distortions, we make use of the distant-observer approximation in the mock galaxy catalogs.  In other words, we add distortions using the peculiar velocity component along a single coordinate axis of the simulation cubes. The linear bias of galaxies in our mock catalogs is approximately $b\sim$ 2.2 and 2.6 for the lower and higher luminosity samples, respectively. This range of bias is roughly consistent with both SDSS LRGs \citep{Marin:2011}, and BOSS galaxies \citep{Parejko:2013, Nuza:2013, Guo:2013}, so results from our mock samples are relevant to both survey data sets.

\subsection{Survey Equivalent Volumes}
\label{ss:volumes}

We wish to estimate the observational constraints from measurements of variance, $S_3$ and $S_4$ using realistic sized surveys.  For this reason, we create subsets of our total simulation volume to match the volumes of the SDSS-II LRG and BOSS samples (in both cases, however, we use the SDSS II LRG mock galaxies described above). SDSS-II has a sky coverage of about $8000 \mathrm{deg}^2$ and the brightest LRG sample can reach a redshift $z \sim 0.45$ \citep{Eisenstein:2001}, which results in a comoving volume of approximately $1 \hgvol$. BOSS covers about $10000 \mathrm{deg}^2$ area and includes galaxies out to $z \sim 0.7$ \citep{Eisenstein:2011}, corresponding to a comoving volume of approximately $4 \hgvol$. Each of the Oriana simulation boxes has a comoving volume of $2.4^3 \hgvol = 13.8 \hgvol $, which we trim to make many realizations of each survey volume. To do this, we cut each box into slices that have volumes equivalent to SDSS-II or BOSS. We leave gaps between subsets to ensure that there are no overlaps between subsets for our density estimates, even with the largest smoothing scale that we employ ($100 \hmpc$). For the $12$ simulations of $3$ non-Gaussian models, this results in $144$ SDSS-II like surveys ($1 \hgvol$) and $36$ BOSS like surveys ($4 \hgvol$) for each model.  We apply the same method to our Gaussian simulations, but since we start with $40$ Gaussian realizations, we end up with $480$ SDSS-II volume subsets and $120$ BOSS volume subsets.  

\section{Results}
\label{s:results}

In each simulation box we estimate densities $\delta_R$ within top-hat smoothing spheres that are arranged on a grid of positions. Since the total volume covered by these spheres will vary with smoothing scale, we add more spheres on small scales and discard some spheres on large scales to ensure that the total volume covered by spheres is always roughly the same on every scale. We then calculate $\Psi_2$ from equation (\ref{Eq:d2p}), and $S_3$ and $S_4$ from equations (\ref{Eq:S3p}) and (\ref{Eq:S4p}). In all the results that follow, we use a set of ten smoothing scales ranging from $10 \hmpc$ to $100 \hmpc$.

\subsection{Dark Matter}
\label{sss:results_dm}

\begin{figure*}[t]
\begin{center}
\includegraphics[scale=0.8]{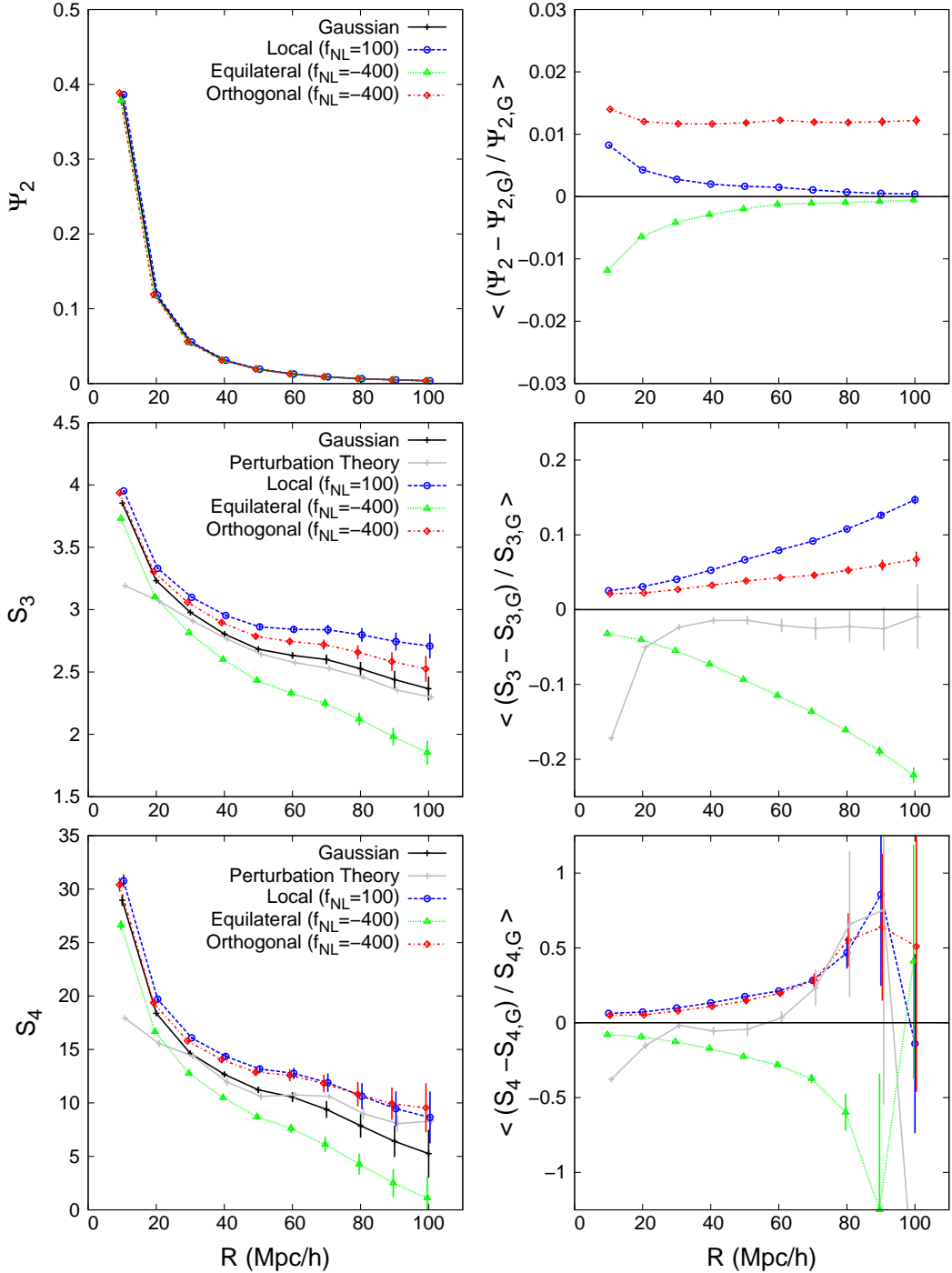}
\caption{
$\Psi_2$ (top left panel), $S_3$ (middle left panel), and $S_4$ (bottom left panel) measurements as a function of smoothing scale for the dark matter distribution in the Gaussian and non-Gaussian simulations. Also shown are the theoretical predictions for $S_3$ and $S_4$ in the Gaussian case from perturbation theory, which are numerically calculated using equations~(\ref{Eq:S3theory}) and (\ref{Eq:S4theory}). Right hand panels show the corresponding residuals of all models with respect to the Gaussian model. In all cases, points show the mean of 12 simulation realizations and error bars show the uncertainty in the mean calculated from their standard deviation. Residuals are likewise calculated separately for each realization and then averaged.
}
\label{Fig:S2S3S4_DM}
\end{center}
\end{figure*}

We first focus on the moments measured from the full dark matter particle distribution.  We use all 12 simulation boxes for each non-Gaussian model, as well as the 12 Gaussian boxes with matching initial phases.  Figure~\ref{Fig:S2S3S4_DM} shows the variance (top panels), skewness (middle panels), and kurtosis (bottom panels) parameters as a function of scale for these different models. The points in the left three panels represent the mean $\Psi_2$, $S_3$, and $S_4$ from the 12 realizations and the error bars show the uncertainty of the mean estimated from their standard deviation. In the case of skewness and kurtosis, we also show the the perturbation theory prediction of the Gaussian model, which we calculate for each realization using equations (\ref{Eq:S3theory}), (\ref{Eq:S4theory}), and (\ref{Eq:gamma}). We evaluate the derivatives in equation (\ref{Eq:gamma}) by spline fitting $\sigma^2$ as a function of smoothing scale and then taking the numerical derivatives. For each realization, we also calculate the residuals between each non-Gaussian model and the Gaussian model, and we show the mean residuals over the 12 realizations along with their errors in the three right panels of Figure~\ref{Fig:S2S3S4_DM}. Since each realization of the non-Gaussian models and the Gaussian model have the same initial phases, the residuals calculated in this way are not sensitive to cosmic variance. 

Let us first focus on results for the variance $\Psi_2$, shown in the top two panels. The variance in the non-Gaussian models is almost identical to that of the Gaussian case. The residuals show that the local and equilateral models have a $\sim 1$\% deviation from the Gaussian variance on the smallest scale we consider, but this deviation vanishes at larger scales. In contrast, the orthogonal model has a roughly constant $\sim 1$\% deviation from the Gaussian variance at all scales. We now move on to the skewness $S_3$, shown in the middle two panels. The residuals clearly show that different non-Gaussian models have different skewness, and the discrepancy increases with scale. For example, the local non-Gaussian model has a skewness that is 3\% higher than the Gaussian model at a scale of $10 \hmpc$, but climbs to 15\% when measured using $100 \hmpc$ smoothing. The difference in sign of the residuals with respect to the Gaussian case is not determined by $\fnl$ alone, e.g. local and orthogonal have positive residuals (despite having opposite signs of $\fnl$) and equilateral has negative residuals (despite having the same $\fnl$ as orthogonal). This is a result of integrating over the non-trivial configuration dependence of the bispectrum in each case (see Eqs.~\ref{Eq:local}$-$\ref{Eq:orthogonal}). Note that though the departure from the Gaussian model grows with scale, so do the skewness error bars.  It is thus not obvious from this result which scales can yield the tightest constraints on models.  We investigate this further below. 
Lastly, we turn to the kurtosis parameter $S_4$, shown in the bottom two panels. The difference between the kurtosis of the Gaussian and non-Gaussian models is clear to see and is actually larger than it was for the skewness, reaching as high as 50\% at large scales. However, the error bars for our kurtosis measurements are substantially larger than they were for the skewness, such that the signal-to-noise of the measurement actually worsens. 

Figure~\ref{Fig:S2S3S4_DM} also shows that the perturbation theory prediction for $S_3$ in the Gaussian case, as given by equation~\ref{Eq:S3theory}, is fairly accurate on scales larger than $30 \hmpc$. We detect a 2\% offset that is consistent with loop corrections in perturbation theory \citep{Scoccimarro:1996, Fosalba:1998}, which are not included in equation~(\ref{Eq:S3theory}). On smaller scales, the accuracy of the prediction drops dramatically, which is expected since perturbation theory breaks down on those scales. The perturbation theory prediction for $S_4$, which we calculate using equation (\ref{Eq:S4theory}), is fairly accurate on scales larger than $30\hmpc$, but fails substantially on smaller scales, as expected. The discrepancy seen for scales larger than $70 \hmpc$ is not statistically significant.

\subsection{Mock Galaxy Catalogs}
\label{sss:results_mocks}

\begin{figure}[t]
\begin{center}
\includegraphics[scale=0.8]{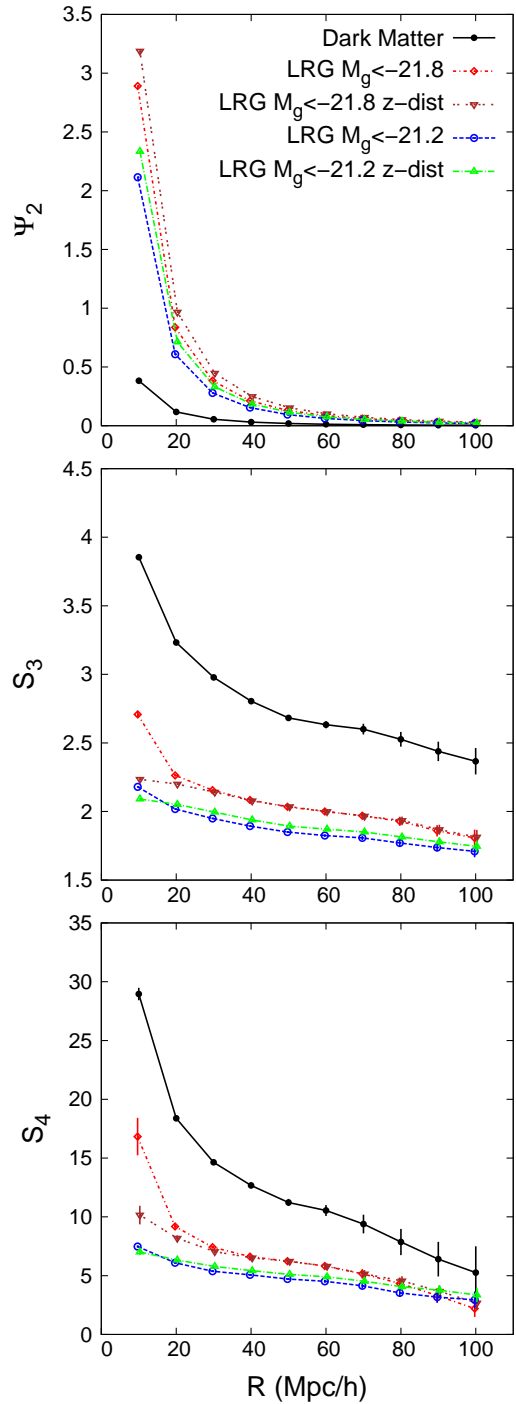}
\caption{
$\Psi_2$ (top panel), $S_3$ (middle panel), and $S_4$ (bottom panel) measurements as a function of smoothing scale on Gaussian simulations for dark matter particles and two mock galaxy catalogs corresponding to SDSS LRGs with $M_g<-21.8$ and $M_g<-21.2$. For each galaxy sample, results are shown both with and without redshift distortions. As in Fig.\ref{Fig:S2S3S4_DM}, points show the mean of 12 simulation realizations and error bars show the uncertainty in the mean.
}
\label{Fig:S2S3S4_DM_Mock_zdist}
\end{center}
\end{figure}

We next investigate the role of galaxy bias on the moments of the density field by measuring them on mock galaxy catalogs instead of the full dark matter distribution.  The two catalogs we use correspond to two SDSS LRG samples with absolute magnitude thresholds of $M_g<-21.8$ and $M_g<-21.2$. We measure $\Psi_2$, $S_3$, and $S_4$ for all the mock catalogs (12 simulation realizations $\times$ 4 sets of initial conditions $\times$ 2 galaxy samples) using the same method we applied to the dark matter particles. Figure~\ref{Fig:S2S3S4_DM_Mock_zdist} shows $\Psi_2$ (top panel) $S_3$ (middle panel), and $S_4$ (bottom panel) measurements on dark matter particles and the two mock galaxy catalogs in the Gaussian case. Mock galaxy results are shown both with and without redshift distortions.

Galaxy bias boosts the variance on all scales, as expected. This is because the variance of the galaxy density field is equal to the variance of the mass field times the linear bias factor squared. Since SDSS LRGs have a bias factor of $\sim 2$, we expect their variance to be roughly four times higher than that of the mass field. Moreover, we expect the more luminous (and thus highly biased) LRG sample to have a higher variance than the lower luminosity sample, which is also clear in the top panel of Figure~\ref{Fig:S2S3S4_DM_Mock_zdist}. Redshift distortions lead to a small increase in the variance on all the scales that we consider. This is because our scales are all in the quasilinear regime where distortions boost the clustering.

\begin{figure*}[t]
\begin{center}
\includegraphics[scale=0.7]{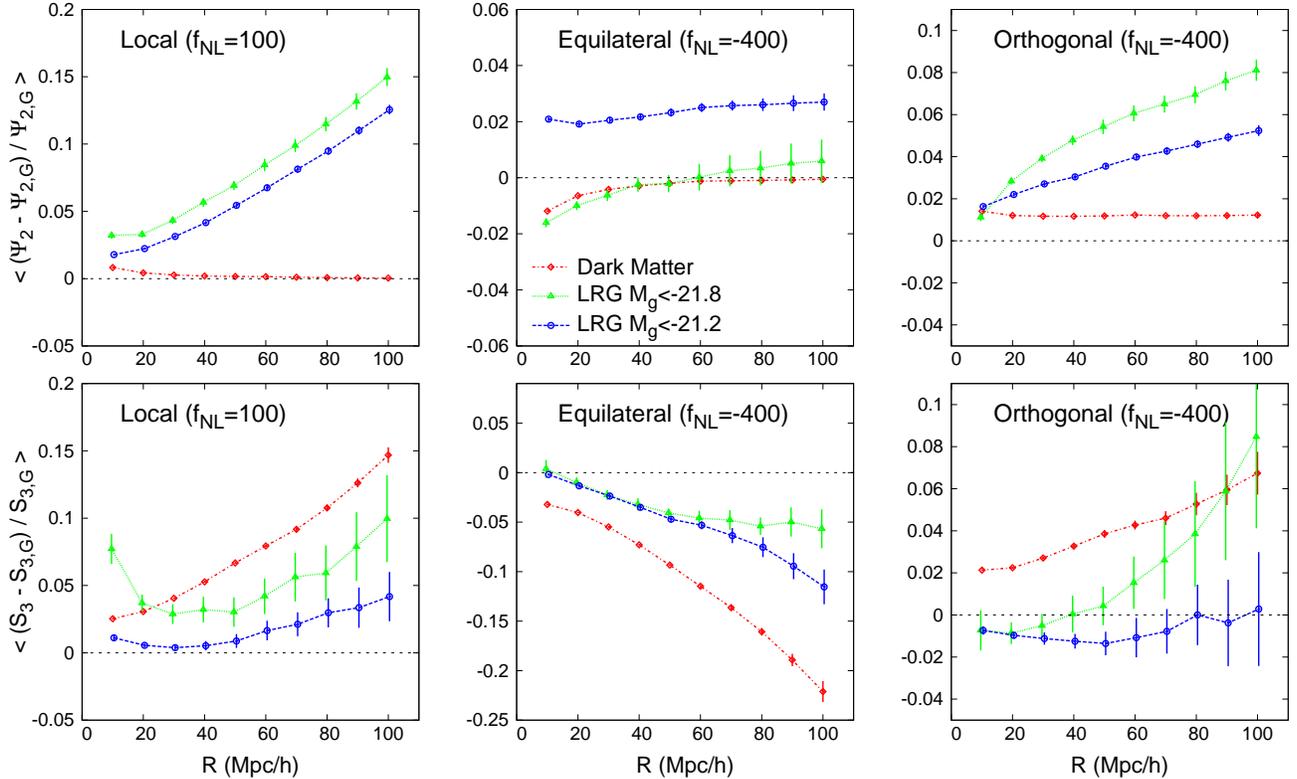}
\caption{
The effect of galaxy bias on the $\Psi_2$ (top panels) and $S_3$ (bottom panels) residuals of non-Gaussian models with respect to the Gaussian model.  Each column of panels shows a different non-Gaussian model and the three different lines show residuals for dark matter (red dot-dashed lines), mock LRGs with $M_g<-21.8$ (green dotted lines) and mock LRGs with $M_g<-21.2$ (blue dashed lines). The residuals are averaged over 12 simulation realizations and error bars show the uncertainty of the mean.
}
\label{Fig:S2S3_residuals_bias}
\end{center}
\end{figure*}

Galaxy bias also has a large effect on $S_3$, decreasing its amplitude by $\sim30-40$\%. This is because the skewness generally scales with the inverse of the linear bias factor, as seen in equation~(\ref{Eq:biasS3}). It is interesting that the more luminous (and highly biased) sample has a higher skewness than the lower bias sample. This is due to the nonlinear quadratic bias term in the same equation, which is larger for the more luminous sample. Redshift distortions generally reduce the skewness on small scales and boost it on large scales; however the exact effect depends on the galaxy sample.  In the more luminous of our two samples redshift distortions do not affect the skewness on scales larger than $30\hmpc$, whereas in the less luminous sample redshift distortions boost the skewness by $\sim$2\% on large scales. Similar results for galaxy bias and redshift distortions hold for the kurtosis $S_4$.

We next focus on the effect of galaxy bias and redshift distortions on non-Gaussian models and, in particular, on the detectability of the models.  In other words, we investigate to what extent bias and redshift distortions affect non-Gaussian models differently from Gaussian models. In this discussion, we only show results for $\Psi_2$ and $S_3$ because $S_4$ is substantially noisier than $S_3$. Figure~\ref{Fig:S2S3_residuals_bias} shows the $\Psi_2$ (top panels) and $S_3$ (bottom panels) residuals between the three non-Gaussian models (each in a different panel) and the Gaussian model for dark matter and the two mock galaxy catalogs. As before, residuals are first calculated for each realization and then averaged. 

\begin{figure*}[t]
\begin{center}
\includegraphics[scale=0.7]{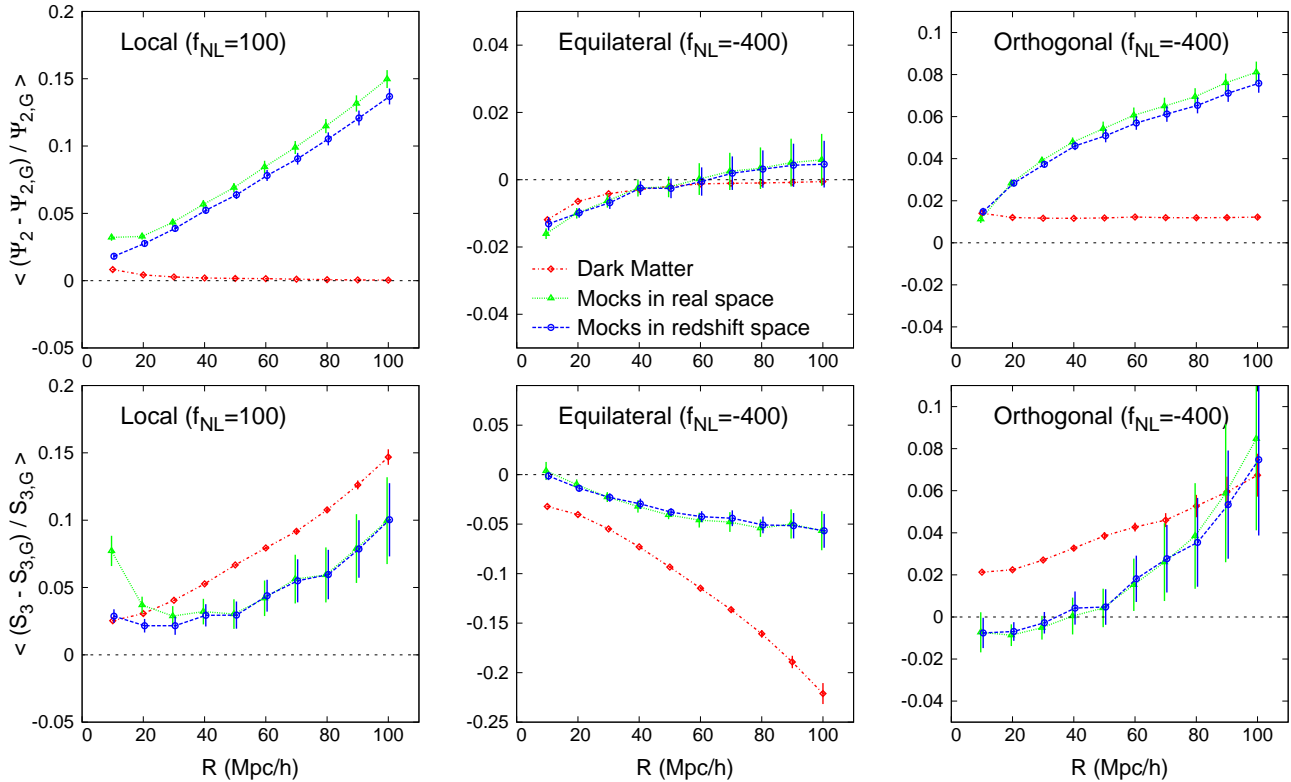}
\caption{
The effect of redshift distortions on the $\Psi_2$ (top panels) and $S_3$ (bottom panels) residuals of non-Gaussian models with respect to the Gaussian model. Each column of panels shows a different non-Gaussian model and the three different lines show residuals for dark matter (red dot-dash lines), mock galaxies in real space (green dot lines) and mock galaxies in redshift space (blue dash lines). The mock galaxies in both cases represent LRGs with $M_g<-21.8$. The residuals are averaged over 12 simulation realizations and error bars show the uncertainty of the mean.
}
\label{Fig:S2S3_residuals_zdist}
\end{center}
\end{figure*}

Galaxy bias has a dramatic effect on the detectability of non-Gaussianity using the variance, particularly for the local case, as expected from power spectrum results \citep{Dalal:2008}. While the deviation from the Gaussian model in the dark matter density field is minimal, it becomes significant in the galaxy mock catalogs. This is because non-Gaussianity leads to corrections in the linear bias factor. In the case of the local non-Gaussian model (top left panel), the fractional difference of the galaxy variance relative to the Gaussian model climbs steadily with scale and reaches as high as 15\% at the largest scale we consider. In addition, the more luminous sample shows a larger deviation than the lower luminosity sample. This behavior is consistent with the bias correction term given by equation~(\ref{Eq:biascorrection}), which shows that the correction grows with both scale and the bias itself. We see similar qualitative behavior for the orthogonal non-Gaussian model (top right panel), though the overall amplitude of the effect is smaller. This is understood from the squeezed limit of the orthogonal template used here, which generates a $1/k$ bias, as discussed after equation~(\ref{Eq:orthogonal}). In the equilateral non-Gaussian model, however, results are different, with the more luminous sample showing a negligible deviation from the Gaussian case, and the less luminous sample showing a $\sim 2$\% higher variance at all scales.

Looking at the skewness, the residuals for the mock galaxies have much larger uncertainties than for the dark matter because the galaxy catalogs have much lower number densities. Nevertheless, it is clear that for all three non-Gaussian models galaxy bias significantly reduces the deviation of the skewness parameter from the Gaussian case. Moreover, this reduction is scale dependent, indicating that scale dependent bias affects the detectability of non-Gaussian models.  The precise relationship between the amount of galaxy bias and the residual skewness is complex and depends on both scale and choice of non-Gaussian model.  For example, in the local and orthogonal non-Gaussian models the more luminous mock galaxy sample yields a larger residual than the lower luminosity mock sample.  However, in the equilateral non-Gaussian model the opposite is true. Though we do not show results for the kurtosis, we find similar results as we did for skewness: galaxy bias generally degrades the discrepancy (and hence detectability) between non-Gaussian and Gaussian models.

Figure~\ref{Fig:S2S3_residuals_zdist} shows the effect of redshift distortions on the $\Psi_2$ (top panels) and $S_3$ (bottom panels) residuals.  Each panel shows a different non-Gaussian model and shows results for the more luminous $M_g<-21.8$ mock galaxy sample.  In almost all cases, the difference between the results on mock galaxy catalogs with and without redshift distortions is negligible. The only exception to this is a $\sim 1$\% deviation for the variance of the local non-Gaussian model. Redshift distortions thus affect $\Psi_2$ and $S_3$ similarly in the Gaussian and non-Gaussian cases, and therefore do not affect the detectability of non-Gaussianity from these measurements.  We find the same qualitative result when investigating the kurtosis.

\subsection{SDSS-II and BOSS Equivalent Volumes}
\label{sss:results_surveys}

We have shown that non-Gaussian initial conditions leave signatures in the skewness of the evolved dark matter density field and that these signatures remain (though diminished) in the galaxy density field as measured in redshift space.  In the case of the variance, non-Gaussian initial conditions leave their strongest signatures in the galaxy density field. However, in most cases we have investigated, the differences from the Gaussian model are fairly small and they tend to be strongest at the largest scales, where cosmic variance errors are also large. In order to quantify whether moments of the density field can be used to constrain non-Gaussian models with measurements from current galaxy surveys, we now use the subsets of our mock catalogs that have volumes equivalent to those of SDSS-II and BOSS, as described in \S\ref{ss:volumes}. 

\begin{figure*}[]
\begin{center}
\includegraphics[scale=0.8]{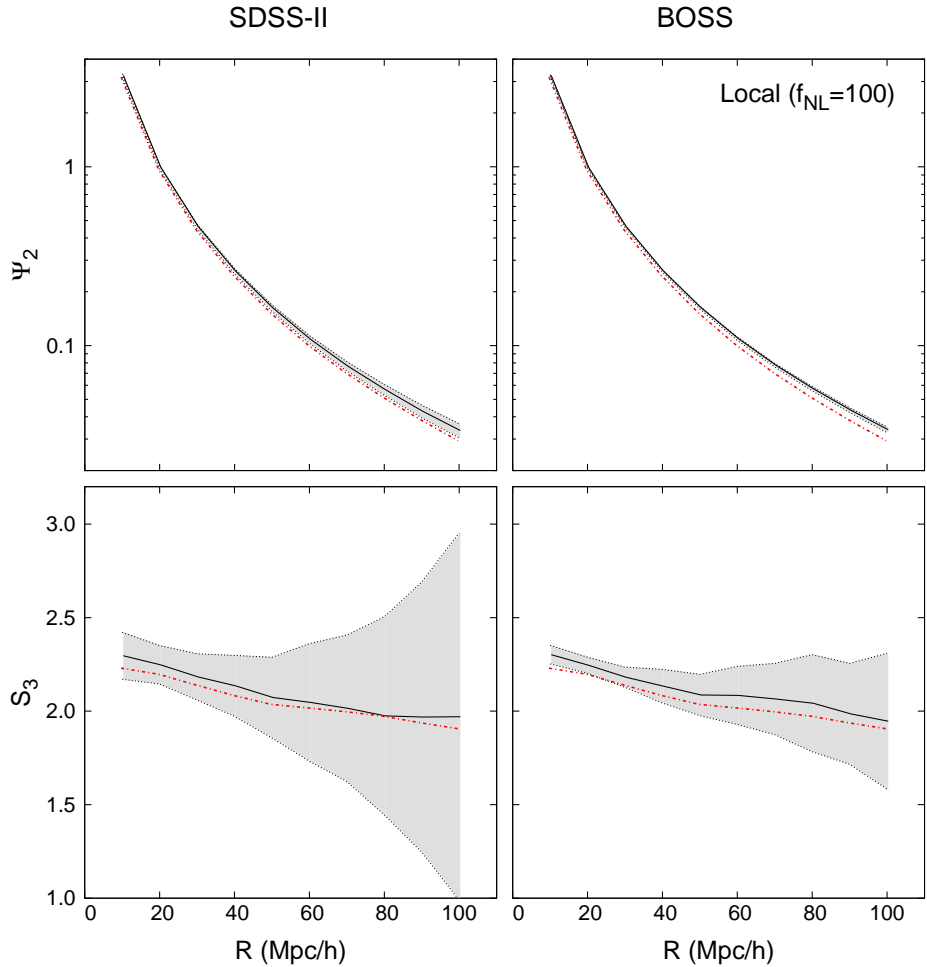}
\caption{
$\Psi_2$ (top panels) and $S_3$ (bottom panels) measurements on SDSS-II (left panels) and BOSS (right panels) equivalent volumes. Each panel shows the mean (solid black curve) and standard deviation (shaded grey region) of the variance or skewness for the local non-Gaussian model, as measured from many independent samples of volume equal to the SDSS-II LRG or BOSS survey. In all cases, the measurements are made on mock LRGs with $M_g<-21.8$ in redshift space.  For comparison, the red dot-dashed curves show the result for the Gaussian case, averaged over all our Gaussian simulations.
}
\label{Fig:S2S3_Mock_demo}
\end{center}
\end{figure*}

Figure~\ref{Fig:S2S3_Mock_demo} shows the distribution of $\Psi_2$ and $S_3$ measurements over these subsets of the local non-Gaussian simulations, for the case of mock LRGs with $M_g<-21.8$ in redshift space. Solid black curves show the mean over all subsets and the shaded grey regions show their standard deviation.  The shaded regions thus span the range of skewness values that we would likely measure from SDSS-II (left panels) or BOSS (right panels) if the local non-Gaussian model studied here correctly described the universe.  For comparison, the red dot-dashed curves show the Gaussian case, which is averaged over all 40 of our Gaussian simulation boxes. The moments of the Gaussian model are thus very well defined and we can ignore their uncertainties. Comparing the red dot-dashed curves with the shaded regions gives us a sense of how well we can discriminate between the local non-Gaussian model and the Gaussian model by measuring $\Psi_2$ or $S_3$ from one of the surveys. Since the skewness of the Gaussian model lies within the $1\sigma$ range of measurements on all scales and for both survey volumes, it is clear that these surveys will not have the power to detect local non-Gaussianity using skewness measurements, even at the unrealistically large $\fnl$ value of our simulations. However, 
the variance of the Gaussian model lies well outside the shaded region - especially for the BOSS equivalent volumes. This means that a measurement of the variance from the BOSS survey could in principle be used to detect local non-Gaussianity with our adopted $\fnl$ value.

\begin{figure*}[]
\begin{center}
\includegraphics[scale=0.7]{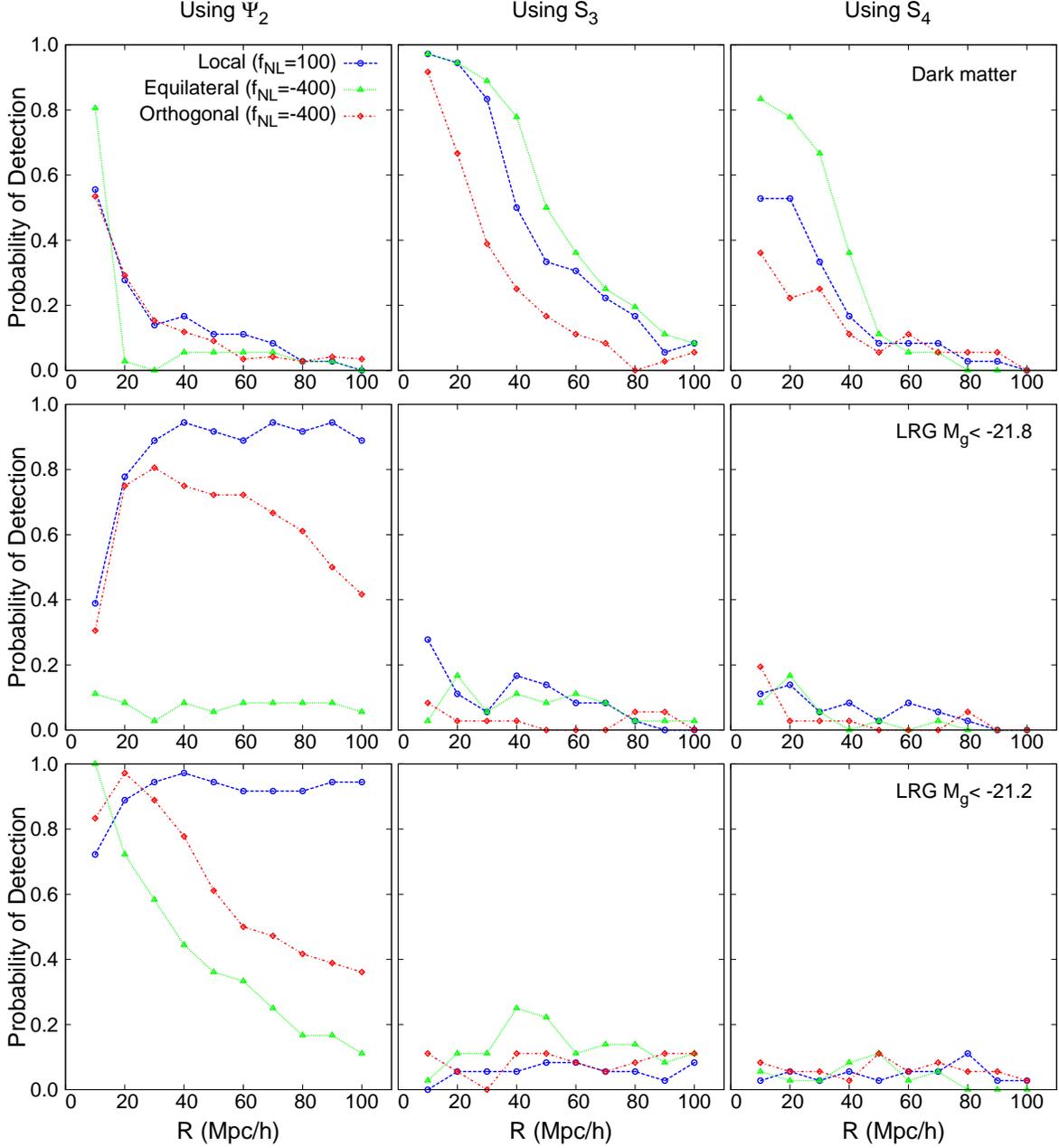}
\caption{
The probability that a measurement of $\Psi_2$ (left panels), $S_3$ (middle panels), or $S_4$ (right panels) in the BOSS galaxy survey can be used to detect a deviation from the Gaussian model at the $2 \sigma$ level. Results are shown for the idealized case of dark matter (top panels), as well as mock LRGs with $M_g<-21.8$ (middle panels) and $M_g<-21.2$ (bottom panels). The three curves in each panel represent the three non-Gaussian models that we explore in this paper: local (blue dashed curves), equilateral (green dotted curves), and orthogonal (red dot-dashed curves). The probabilities are given by the percentage of BOSS survey equivalent volumes that have a $\chi^2$ value higher than the value corresponding to the $2 \sigma$ level, where each $\chi^2$ value is calculated from comparing the measurement from a single non-Gaussian sample volume to the mean of all Gaussian realizations. See \S \ref{sss:results_surveys} for details.
}
\label{Fig:S2S3S4_BOSS_chisquare}
\end{center}
\end{figure*}

To quantify this result, we calculate the likelihood that a given survey will detect the departure from primordial Gaussianity using a $\Psi_2$, $S_3$, or $S_4$ measurement on each scale. For each non-Gaussian model (e.g., local model), mock galaxy luminosity (e.g., LRGs with $M_g<-21.8$), survey volume (e.g., BOSS), choice of moment (e.g., skewness), and scale (e.g., $20\hmpc$), we calculate the $\chi^2$ value between the measurement of each non-Gaussian subset and the ``true'' Gaussian measurement. For example, in the case of the skewness, this is
\beq
\label{Eq:chi2}
\chi_i^2 = \frac{(S_{3NG,i}-S_{3G})^2}{(\sigma_{S_{3G}})^2},
\eeq
where $S_{3NG,i}$ is the skewness of the $i$th non-Gaussian subset (we have 144 subsets for SDSS-II and 36 for BOSS), $S_{3G}$ is the ``true'' skewness of the Gaussian model measured from all 40 Gaussian simulation boxes, and $\sigma_{S_{3G}}$ is the standard deviation of skewness values measured from all the Gaussian subsets (we have 480 subsets for SDSS-II and 120 for BOSS). Note that $\sigma_{S_{3G}}$ is not the standard deviation of the non-Gaussian subsets as shown in Figure~\ref{Fig:S2S3_Mock_demo}. We choose to use the Gaussian subsets because we have many more Gaussian simulations and so the standard deviation can be more accurately estimated than from the non-Gaussian subsets.  Moreover, the standard deviation is dominated by shot noise and cosmic variance and we do not expect it to vary significantly between the Gaussian and non-Gaussian models. Once we have a $\chi^2$ value for each realization of a survey volume, we estimate the fraction of these values that exceed the value corresponding to a $2\sigma$ detection.  For one degree of freedom, this value is 4.  The fraction of realizations that have $\chi^2>4$ is thus the probability that a measurement from such a survey would be able to provide $2\sigma$ evidence for $\fnl$.

\begin{figure*}[]
\begin{center}
\includegraphics[scale=0.7]{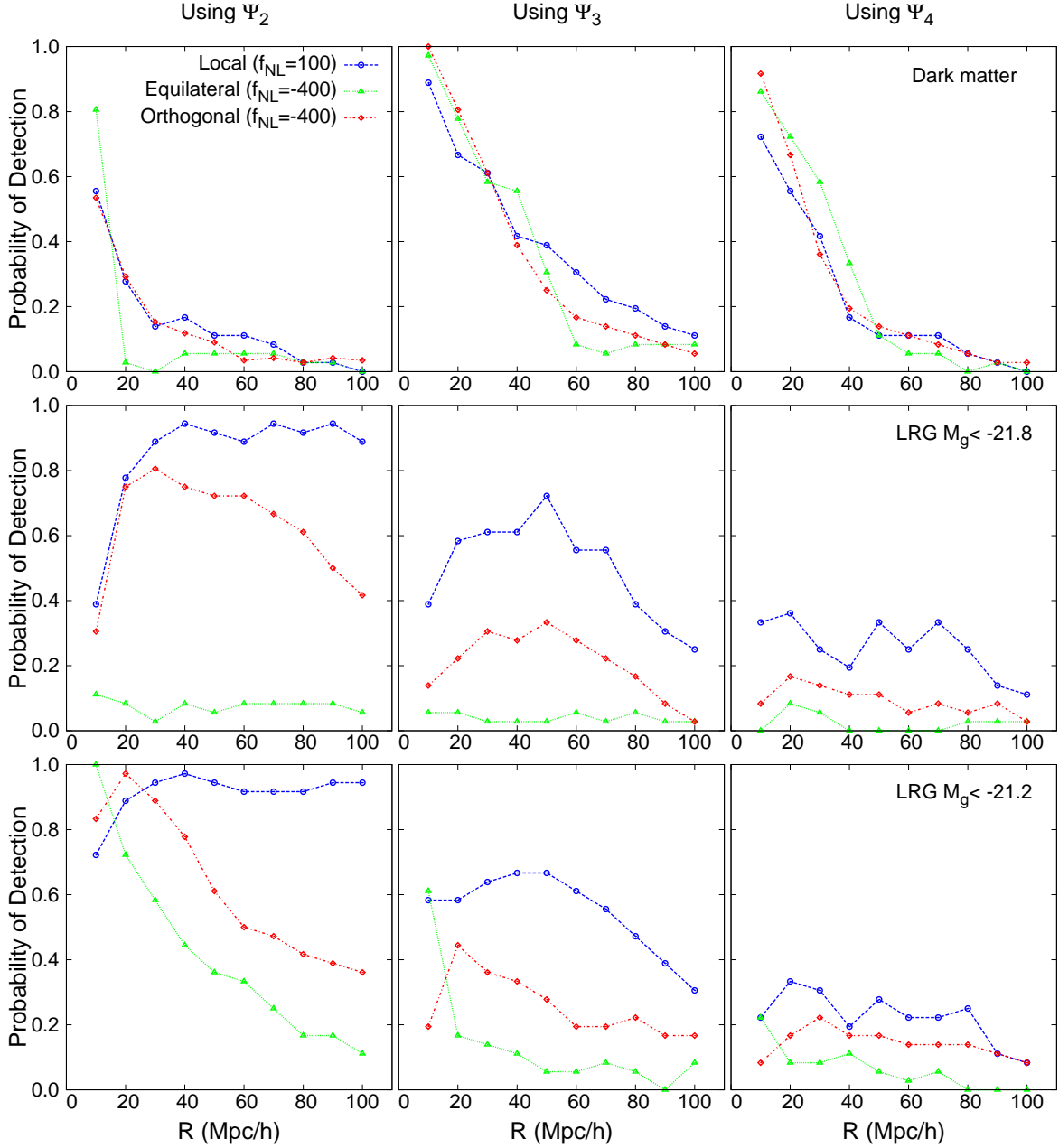}
\caption{
Similar in every respect to Fig~\ref{Fig:S2S3S4_BOSS_chisquare}, except that the skewness and kurtosis are replaced by the third and fourth moments, $\Psi_3$ and $\Psi_4$.
}
\label{Fig:Psi2Psi3Psi4_BOSS_chisquare}
\end{center}
\end{figure*}

Figure~\ref{Fig:S2S3S4_BOSS_chisquare} shows these probabilities as a function of smoothing scale for a BOSS equivalent volume. Each panel shows results for one combination of density field moment and density field tracer and the three curves in each panel show results for the three non-Gaussian models. The results are somewhat noisy because we only have a limited number of BOSS survey volume subsets, but the main conclusions are clear.  Skewness measurements at small scales ($10-20\hmpc$) using the full dark matter distribution would be able to provide evidence for $\fnl$ in the BOSS survey.  However, the probabilities drop dramatically when galaxies are used instead of dark matter.  The probability of detecting non-Gaussianity by measuring the galaxy skewness in BOSS is at best $\sim 25$\%, and closer to 10\% in most cases, even for our unrealistically high $\fnl$ models. Kurtosis measurements are noisier and thus even less likely to yield a detection of non-Gaussianity. Figure~\ref{Fig:S2S3S4_BOSS_chisquare} shows that the likelihood of detecting our non-Gaussian models with a kurtosis measurement of the BOSS galaxy density field is below the 10\% level for almost all galaxy samples, scales and non-Gaussian models.

The story is different, however, in the case of the variance. Figure~\ref{Fig:S2S3S4_BOSS_chisquare} shows that measurements of $\Psi_2$ from a galaxy survey like BOSS would in principle be able to detect our non-Gaussian models if they correctly described the universe. The optimal galaxy sample and smoothing scale depends on the specific non-Gaussian model. For example, if the universe were described by our local non-Gaussian model, we would have a more than 90\% chance of detecting the departure from Gaussianity at the 2$\sigma$ level with either of our galaxy samples and on any scale larger than $30\hmpc$. In the case of the orthogonal model, we would need to measure the variance on scales of $20-30\hmpc$ with either sample. On the other hand, the equilateral model would only be detected with a small scale measurement at $10\hmpc$, and only using our less luminous LRG sample. The more luminous sample would yield no detection at all.

The success of the variance raises the question of whether the higher order moments would provide more constraining power if they were not normalized by the variance. In other words, what happens when we use $\Psi_3$ and $\Psi_4$, instead of $S_3$ and $S_4$? We show results for this in Figure~\ref{Fig:Psi2Psi3Psi4_BOSS_chisquare}, which is exactly the same as Figure~\ref{Fig:S2S3S4_BOSS_chisquare} in every respect except that $S_3$ and $S_4$ are replaced by $\Psi_3$ and $\Psi_4$. The figure shows that the raw moments perform much better than their normalized versions when galaxy samples are used. However, they do not perform as well as the variance, and they are expected to be more covariant with $\Psi_2$ than $S_3$ and $S_4$ given what is known from the bispectrum versus reduced bispectrum \citep{Sefusatti:2006}. For all non-gaussian models, $\Psi_2$ shows the most promise, followed by $\Psi_3$ and then $\Psi_4$. $S_3$ and $S_4$ come last, showing the least constraining power in detecting $\fnl$. We thus conclude that the constraining power of the higher order moments essentially comes from the variance and that any additional information that exists in the higher moments provides minimal constraint.

In summary, the N-body simulations clearly show that Gaussian and different non-Gaussian initial conditions lead to different moments in the evolved density field. However, the probability of detecting this inconsistency with the Gaussian model by measuring the moments on a galaxy survey like SDSS-II or BOSS is in most cases low, even for the unrealistically large amplitude non-Gaussian models that we consider here. The variance of the galaxy density field is the only measurement that could in principle detect evidence for our $\fnl$ models.

\subsection{Scaling Down to Realistic $\fnl$ Values}
\label{ss:scaling_fnl}

It would be useful to know how much survey volume is needed to reliably detect a deviation from the Gaussian model using the skewness. We estimate the volume needed by rescaling the standard deviation of skewness in equation~(\ref{Eq:chi2}). On large scales, the standard deviation of the skewness is dominated by cosmic variance and we simply assume that it scales as $1/\sqrt{V}$, where $V$ is the survey volume. We rescale the standard deviation of the skewness in this way and recalculate the probability that a skewness measurement from a survey with a given volume can provide $2\sigma$ evidence for $\fnl$, following the methodology described in \S~\ref{sss:results_surveys}. We find that if our local non-Gaussian model ($\fnl=100$) correctly described the universe, we would need a survey volume that is 1.5 times the BOSS volume to have a 50\% likelihood of detecting non-Gaussianity by measuring the skewness of LRGs with $M_g<-21.8$ on a $10\hmpc$ scale. If instead our equilateral non-Gaussian model ($\fnl=-400$) correctly described the universe, we would need a survey volume that is 2.3 times the BOSS volume to have a 50\% likelihood of detecting non-Gaussianity by measuring the skewness of LRGs with $M_g<-21.2$ on a $40\hmpc$ scale. Note that while we have considered here a single scale $R$ in deriving constraints, including more scales in the analysis is not expected to qualitatively change the conclusions, since different smoothing scales are significantly covariant.

These results apply to the specific non-Gaussian models that we consider in this paper, which have large non-Gaussian amplitudes compared to what is allowed from recent Planck constraints. Any realistic departures from Gaussianity (if they exist) in the universe are thus far smaller than what we have studied and will require even larger survey volumes to detect them. We can estimate how much larger a survey volume is needed for a realistic model, by scaling the volume by $1/\fnl^2$, since the primordial skewness is directly proportional to $\fnl$ (keeping shot noise and galaxy bias fixed). For example, adopting the Planck $1\sigma$ constraints, a local non-Gaussian model with $\fnl=6$ would require a volume that is $\sim 280$ times larger than the one needed to detect our $\fnl=100$ model. This is much larger than the future Euclid survey, which will have approximately 30 times more volume than BOSS. An equilateral non-Gaussian model with $\fnl=-75$ would require a volume that is approximately 30 times larger than the one needed to detect our $\fnl=-400$ model. This is also larger than what Euclid will provide. We can therefore conclude that skewness and kurtosis measurements are never likely to yield a detection of primordial non-Gaussianity of inflationary type. In principle, there could be other types of non-Gaussianities for which this conclusion may not hold, e.g. non-Gaussianity that is the result of nonlinearities in the density rather than the potential \citep{Scherrer:1995,Verde:2001}. 

In the case of the variance, we can also find the survey volume that would be needed to detect non-Gaussian models with realistic $\fnl$ amplitudes. First let us find the volume that would be needed to yield a 50\% likelihood of detecting our unrealistically high $\fnl$ models. We can do that by scaling the denominator in equation~(\ref{Eq:chi2}) when rewritten for $\Psi_2$ (i.e., the standard deviation of the variance) by $1/\sqrt{V}$.  Then we can recalculate the probability that a variance measurement from a survey with a given volume can provide $2\sigma$ evidence for $\fnl$. We find that we would need a survey volume that is 0.37 times that of BOSS to have a 50\% likelihood of detecting our local non-Gaussian model by measuring the variance of LRGs with $M_g<-21.2$ on a $40\hmpc$ scale. In the case of the equilateral model, we would need a volume that is 0.33 times that of BOSS when using a variance measurement on the same galaxy sample on a $10\hmpc$ scale. Now we can estimate what survey volumes are necessary to detect more realistic $\fnl$ amplitudes. As before, detecting a local non-Gaussian model with $\fnl=6$ requires 280 times more volume, while detecting an equilateral non-Gaussian model with $\fnl=-75$ requires 30 times more volume. For our best case sets of galaxy sample and scales, this translates to survey volumes that are 100 and 10 times larger than BOSS, respectively. The first is larger than any future survey, but the second will be achieved by Euclid.

\subsection{Comparison With Existing Measurements}

\begin{figure*}[t]
\begin{center}
\includegraphics[scale=0.7]{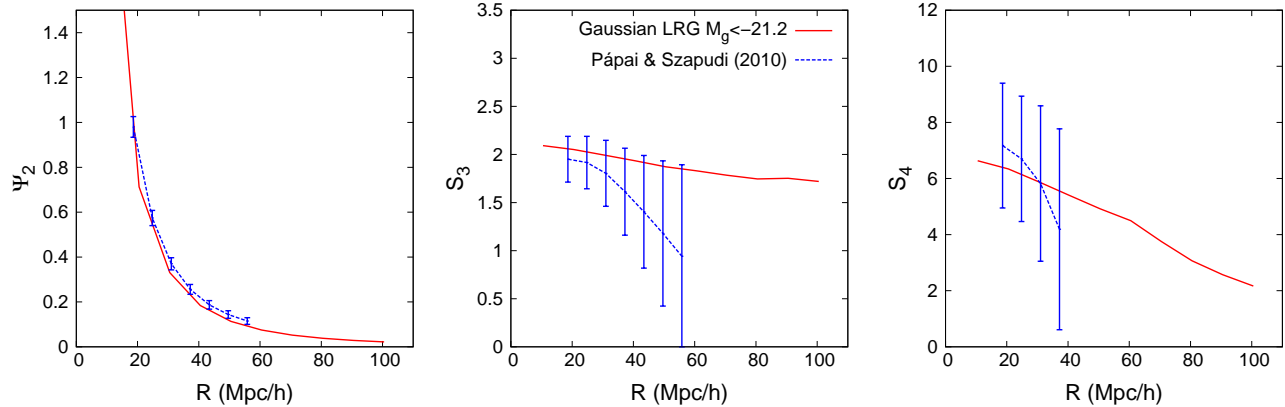}
\caption{
A comparison of the \cite{P'apai:2010} measurements of $S_3$ (top panel) and $S_4$ (bottom panel) from SDSS-II data with measurements from our Gaussian simulations. The dashed blue curves show the \cite{P'apai:2010} measurements on a SDSS-II LRG sample using CIC smoothing in cubic cells, along with their $1\sigma$ uncertainties. The solid red curves show measurements from our Gaussian simulations, using mock LRGs with $M_g<-21.2$ in redshift space. The Gaussian results represent an average over 120 SDSS-II equivalent volumes and the error bars show their standard deviation. Note that we have shifted the scales of the \cite{P'apai:2010} measure ments to account for the different definition of smoothing filter between their and our work.
}
\label{Fig:SDSS_data}
\end{center}
\end{figure*}

We have demonstrated that measurements of moments of the galaxy density field in existing survey data are unlikely to provide meaningful constraints on primordial non-Gaussianity anytime soon. Nevertheless, these measurements are very easy to make and it is thus still worth making a quick measurement of variance, skewness, and kurtosis in the SDSS-II and BOSS survey data and comparing it to the Gaussian model.

\cite{P'apai:2010} measured $\Psi_2$, $S_3$ and $S_4$ in the SDSS-II DR7 spectroscopic LRG sample. They selected their LRG sample with k-corrected absolute magnitudes between -22.3 and -24.3 in the $r$ band, which is close to our $M_g<-21.2$ sample. Their measurements range from $30-90\hmpc$. However, they used a CIC method with cubic cells so their scales roughly correspond to 1.6 times our scales, i.e. the range of scales in their work corresponds to the range $19-56\hmpc$ in our work. In Figure~\ref{Fig:SDSS_data}, we plot their measurements along with their $1\sigma$ uncertainties, compared with our measurements from our Gaussian model using mock LRGs with $M_g<-21.2$ in redshift space. Our measurements represent the mean of 120 mocks with SDSS-II equivalent volumes. We should be cautious about making a direct comparison between these two results because there are differences in the methods used, including the choice of smoothing filter, estimator and sample selection. Nevertheless, we see that the \cite{P'apai:2010} measurements are consistent with our Gaussian model. Their variance measurements are slightly larger than ours and their skewness measurements are lower than ours, but these differences are not significant given their error bars. 

\section{Summary and Discussion}
\label{s:discussion}

In this paper, we have measured the variance $\Psi_2$, skewness parameter $S_3$, and kurtosis parameter $S_4$ on N-body simulations that are seeded with local ($\fnl=100$), equilateral ($\fnl=-400$), and orthogonal ($\fnl=-400$) non-Gaussian initial conditions, as well as with Gaussian initial conditions. We have made measurements on the evolved dark matter density field and on two different sets of mock galaxy catalogs that were designed to simulate two different luminosity samples. Finally, we have investigated the detectability of non-Gaussianity for different galaxy survey volumes. Our main conclusions are as follows.

\begin{itemize}

\item Simulations seeded with Gaussian and different non-Gaussian initial conditions show different variance, skewness, and kurtosis in the evolved density field. The differences are clear in both the dark matter distribution and mock galaxy catalogs.

\item Galaxy bias, for the LRG-type galaxies that we consider, significantly reduces the detectability of primordial non-Gaussianity using skewness and kurtosis measurements, but dramatically increases the detectability using measurements of the variance. Since different non-Gaussian models provide different scale-dependent bias corrections, the deviation of non-Gaussian models from the Gaussian case depends on the amount of bias and the scale, as well as the nature of the non-Gaussian model.

\item Redshift distortions shift the variance, skewness, and kurtosis in the same way for Gaussian and non-Gaussian initial conditions. As a result, they do not affect the detectability of primordial non-Gaussianity.

\item Skewness and kurtosis measurements made in current galaxy survey volumes will not have sufficient signal-to-noise to detect primordial non-Gaussianity. The likelihood of finding $2\sigma$ evidence for $\fnl$ by making a skewness measurement in a volume equivalent to the BOSS survey is less than $\sim 25$\% for the galaxy samples and scales and $\fnl$ values we consider. Kurtosis measurements provide even worse constraining power. Measurements of the galaxy variance however, have a high probability of detecting our $\fnl$ values in a volume equivalent to BOSS.

\item The unnormalized higher order moments $\Psi_3$ and $\Psi_4$ provide more constraining power than their normalized versions $S_3$ and $S_4$. However, these moments do not perform as well as the variance, and they are expected to be more covariant with $\Psi_2$.

\item Using simple arguments to scale our results to more realistic $\fnl$ values (for example, $\fnl=6$ for the local model and $\fnl=-75$ for the equilateral model), we find that skewness and kurtosis measurements will likely never have sufficient signal-to-noise to detect non-Gaussianity of inflationary type, since the required survey volumes exceed those of the largest planned future surveys. Measurements of the galaxy variance, however, should be able to probe interesting values of $\fnl$ for some non-Gaussian models in a survey like Euclid.
\end{itemize}

These results are not surprising because the skewness and kurtosis only contain reduced information about the density field. They are not nearly as sensitive as the bispectrum and trispectrum when used as a probe of primordial non-Gaussianity. On the other hand, the variance contains very similar information to the power spectrum, which many studies have shown will be able to provide competitive constraints on non-Gaussianity \citep[e.g.,][]{Giannantonio:2012}.
Measurements of the skewness and kurtosis from larger future redshift surveys, such as eBOSS, DESI, and Euclid will have much larger signal-to-noise and will provide tighter constraints on non-Gaussian models. However, as we discussed above, these constraints will not be competitive with already existing constraints from Planck. Only the bispectrum and trispectrum have sufficient constraining power to have a chance at detecting primordial non-Gaussianity. Their higher constraining power results from the shape dependence of these correlators that is lost when integrating it out with spherical top-hat filters to get the skewness and kurtosis parameters. To take advantage of such dependencies, however, nontrivial effects due to bias and redshift-space distortions must be fully accounted for. We hope to report on this soon.

\section{acknowledgments}

Q.M. and A.A.B. were supported by the National Science Foundation (NSF) through NSF CAREER grant AST-1151650. R.J.S. was supported in part by the Department of Energy (DE-FG05-85ER40226). Q.M., A.A.B. and R.J.S. were also supported by a Discovery grant from Vanderbilt University. R.S. is supported in part by grants NSF AST-1109432 and NASA NNA10A171G.
The simulations used in this paper were produced by the LasDamas project (http://lss.phy.vanderbilt.edu/lasdamas/); we thank NSF XSEDE for providing the computational resources for LasDamas. 

\bibliographystyle{apj}
\bibliography{Astro}

\begin{thebibliography}{82}
\expandafter\ifx\csname natexlab\endcsname\relax\def\natexlab#1{#1}\fi

\bibitem[{{Afshordi} \& {Tolley}(2008)}]{Afshordi:2008}
{Afshordi}, N., \& {Tolley}, A.~J. 2008, \prd, 78, 123507

\bibitem[{{Babich} {et~al.}(2004){Babich}, {Creminelli}, \&
  {Zaldarriaga}}]{Babich:2004}
{Babich}, D., {Creminelli}, P., \& {Zaldarriaga}, M. 2004, J. Cosmol.
  Astropart. Phys., 8, 9

\bibitem[{{Baldauf} {et~al.}(2011){Baldauf}, {Seljak}, \&
  {Senatore}}]{Baldauf:2011}
{Baldauf}, T., {Seljak}, U., \& {Senatore}, L. 2011, J. Cosmol. Astropart.
  Phys., 4, 6

\bibitem[{{Bartolo} {et~al.}(2004){Bartolo}, {Komatsu}, {Matarrese}, \&
  {Riotto}}]{Bartolo:2004}
{Bartolo}, N., {Komatsu}, E., {Matarrese}, S., \& {Riotto}, A. 2004, \physrep,
  402, 103

\bibitem[{{Bennett} {et~al.}(2013){Bennett}, {Larson}, {Weiland}, {Jarosik},
  {Hinshaw}, {Odegard}, {Smith}, {Hill}, {Gold}, {Halpern}, {Komatsu}, {Nolta},
  {Page}, {Spergel}, {Wollack}, {Dunkley}, {Kogut}, {Limon}, {Meyer}, {Tucker},
  \& {Wright}}]{Bennett:2013}
{Bennett}, C.~L., {et~al.} 2013, \apjs, 208, 20

\bibitem[{{Benoist} {et~al.}(1999){Benoist}, {Cappi}, {da Costa},
  {Maurogordato}, {Bouchet}, \& {Schaeffer}}]{Benoist:1999}
{Benoist}, C., {Cappi}, A., {da Costa}, L.~N., {Maurogordato}, S., {Bouchet},
  F.~R., \& {Schaeffer}, R. 1999, \apj, 514, 563

\bibitem[{{Berlind} \& {Weinberg}(2002)}]{Berlind:2002}
{Berlind}, A.~A., \& {Weinberg}, D.~H. 2002, \apj, 575, 587

\bibitem[{{Bernardeau}(1992)}]{Bernardeau:1992}
{Bernardeau}, F. 1992, \apj, 392, 1

\bibitem[{{Bernardeau}(1994)}]{Bernardeau:1994}
---. 1994, \apj, 433, 1

\bibitem[{{Bernardeau} {et~al.}(2002){Bernardeau}, {Colombi}, {Gazta{\~n}aga},
  \& {Scoccimarro}}]{Bernardeau:2002}
{Bernardeau}, F., {Colombi}, S., {Gazta{\~n}aga}, E., \& {Scoccimarro}, R.
  2002, \physrep, 367, 1

\bibitem[{{Bouchet} {et~al.}(1992){Bouchet}, {Juszkiewicz}, {Colombi}, \&
  {Pellat}}]{Bouchet:1992}
{Bouchet}, F.~R., {Juszkiewicz}, R., {Colombi}, S., \& {Pellat}, R. 1992,
  \apjl, 394, L5

\bibitem[{{Bouchet} {et~al.}(1993){Bouchet}, {Strauss}, {Davis}, {Fisher},
  {Yahil}, \& {Huchra}}]{Bouchet:1993}
{Bouchet}, F.~R., {Strauss}, M.~A., {Davis}, M., {Fisher}, K.~B., {Yahil}, A.,
  \& {Huchra}, J.~P. 1993, \apj, 417, 36

\bibitem[{{Chen}(2010)}]{Chen:2010}
{Chen}, X. 2010, Advances in Astronomy, 2010

\bibitem[{{Chodorowski} \& {Bouchet}(1996)}]{Chodorowski:1996}
{Chodorowski}, M.~J., \& {Bouchet}, F.~R. 1996, \mnras, 279, 557

\bibitem[{{Coles} \& {Frenk}(1991)}]{Coles:1991}
{Coles}, P., \& {Frenk}, C.~S. 1991, \mnras, 253, 727

\bibitem[{{Coles} {et~al.}(1993){Coles}, {Moscardini}, {Lucchin}, {Matarrese},
  \& {Messina}}]{Coles:1993}
{Coles}, P., {Moscardini}, L., {Lucchin}, F., {Matarrese}, S., \& {Messina}, A.
  1993, \mnras, 264, 749

\bibitem[{{Creminelli} {et~al.}(2006){Creminelli}, {Nicolis}, {Senatore},
  {Tegmark}, \& {Zaldarriaga}}]{Creminelli:2006}
{Creminelli}, P., {Nicolis}, A., {Senatore}, L., {Tegmark}, M., \&
  {Zaldarriaga}, M. 2006, J. Cosmol. Astropart. Phys., 5, 4

\bibitem[{{Crocce} {et~al.}(2006){Crocce}, {Pueblas}, \&
  {Scoccimarro}}]{Crocce:2006}
{Crocce}, M., {Pueblas}, S., \& {Scoccimarro}, R. 2006, \mnras, 373, 369

\bibitem[{{Croton} {et~al.}(2004){Croton}, {Gazta{\~n}aga}, {Baugh}, {Norberg},
  {Colless}, {Baldry}, {Bland-Hawthorn}, {Bridges}, {Cannon}, {Cole},
  {Collins}, {Couch}, {Dalton}, {De Propris}, {Driver}, {Efstathiou}, {Ellis},
  {Frenk}, {Glazebrook}, {Jackson}, {Lahav}, {Lewis}, {Lumsden}, {Maddox},
  {Madgwick}, {Peacock}, {Peterson}, {Sutherland}, \& {Taylor}}]{Croton:2004}
{Croton}, D.~J., {et~al.} 2004, \mnras, 352, 1232

\bibitem[{{Dalal} {et~al.}(2008){Dalal}, {Dor{\'e}}, {Huterer}, \&
  {Shirokov}}]{Dalal:2008}
{Dalal}, N., {Dor{\'e}}, O., {Huterer}, D., \& {Shirokov}, A. 2008, \prd, 77,
  123514

\bibitem[{{Dawson} {et~al.}(2013){Dawson}, {Schlegel}, {Ahn}, {Anderson},
  {Aubourg}, {Bailey}, {Barkhouser}, {Bautista}, {Beifiori}, {Berlind},
  {Bhardwaj}, {Bizyaev}, {Blake}, {Blanton}, {Blomqvist}, {Bolton}, {Borde},
  {Bovy}, {Brandt}, {Brewington}, {Brinkmann}, {Brown}, {Brownstein}, {Bundy},
  {Busca}, {Carithers}, {Carnero}, {Carr}, {Chen}, {Comparat}, {Connolly},
  {Cope}, {Croft}, {Cuesta}, {da Costa}, {Davenport}, {Delubac}, {de Putter},
  {Dhital}, {Ealet}, {Ebelke}, {Eisenstein}, {Escoffier}, {Fan}, {Filiz Ak},
  {Finley}, {Font-Ribera}, {G{\'e}nova-Santos}, {Gunn}, {Guo}, {Haggard},
  {Hall}, {Hamilton}, {Harris}, {Harris}, {Ho}, {Hogg}, {Holder}, {Honscheid},
  {Huehnerhoff}, {Jordan}, {Jordan}, {Kauffmann}, {Kazin}, {Kirkby}, {Klaene},
  {Kneib}, {Le Goff}, {Lee}, {Long}, {Loomis}, {Lundgren}, {Lupton}, {Maia},
  {Makler}, {Malanushenko}, {Malanushenko}, {Mandelbaum}, {Manera}, {Maraston},
  {Margala}, {Masters}, {McBride}, {McDonald}, {McGreer}, {McMahon}, {Mena},
  {Miralda-Escud{\'e}}, {Montero-Dorta}, {Montesano}, {Muna}, {Myers},
  {Naugle}, {Nichol}, {Noterdaeme}, {Nuza}, {Olmstead}, {Oravetz}, {Oravetz},
  {Owen}, {Padmanabhan}, {Palanque-Delabrouille}, {Pan}, {Parejko},
  {P{\^a}ris}, {Percival}, {P{\'e}rez-Fournon}, {P{\'e}rez-R{\`a}fols},
  {Petitjean}, {Pfaffenberger}, {Pforr}, {Pieri}, {Prada}, {Price-Whelan},
  {Raddick}, {Rebolo}, {Rich}, {Richards}, {Rockosi}, {Roe}, {Ross}, {Ross},
  {Rossi}, {Rubi{\~n}o-Martin}, {Samushia}, {S{\'a}nchez}, {Sayres}, {Schmidt},
  {Schneider}, {Sc{\'o}ccola}, {Seo}, {Shelden}, {Sheldon}, {Shen}, {Shu},
  {Slosar}, {Smee}, {Snedden}, {Stauffer}, {Steele}, {Strauss}, {Streblyanska},
  {Suzuki}, {Swanson}, {Tal}, {Tanaka}, {Thomas}, {Tinker}, {Tojeiro},
  {Tremonti}, {Vargas Maga{\~n}a}, {Verde}, {Viel}, {Wake}, {Watson}, {Weaver},
  {Weinberg}, {Weiner}, {West}, {White}, {Wood-Vasey}, {Yeche}, {Zehavi},
  {Zhao}, \& {Zheng}}]{Dawson:2013}
{Dawson}, K.~S., {et~al.} 2013, \aj, 145, 10

\bibitem[{{Eisenstein} {et~al.}(2001){Eisenstein}, {Annis}, {Gunn}, {Szalay},
  {Connolly}, {Nichol}, {Bahcall}, {Bernardi}, {Burles}, {Castander},
  {Fukugita}, {Hogg}, {Ivezi{\'c}}, {Knapp}, {Lupton}, {Narayanan}, {Postman},
  {Reichart}, {Richmond}, {Schneider}, {Schlegel}, {Strauss}, {SubbaRao},
  {Tucker}, {Vanden Berk}, {Vogeley}, {Weinberg}, \& {Yanny}}]{Eisenstein:2001}
{Eisenstein}, D.~J., {et~al.} 2001, \aj, 122, 2267

\bibitem[{Eisenstein {et~al.}(2011)Eisenstein, Weinberg, Agol, Aihara, Prieto,
  Anderson, Arns, Aubourg, Bailey, Balbinot, Barkhouser, Beers, Berlind,
  Bickerton, Bizyaev, Blanton, Bochanski, Bolton, Bosman, Bovy, Brewington,
  Brandt, Breslauer, Brinkmann, Brown, Brownstein, Burger, Busca, Campbell,
  Cargile, Carithers, Carlberg, Carr, Chen, Chiappini, Comparat, Connolly,
  Cortes, Croft, da~Costa, Cunha, Davenport, Dawson, Lee, de~Mello, de~Simoni,
  Dean, Dhital, Ealet, Ebelke, Edmondson, Eiting, Escoffier, Esposito, Evans,
  Fan, Castella, Ferreira, Fitzgerald, Fleming, Font-Ribera, Ford, Frinchaboy,
  Perez, Gaudi, Ge, Ghezzi, Gillespie, Gilmore, Girardi, Gott, Gould, Grebel,
  Gunn, Hamilton, Harding, Harris, Hawley, Hearty, Hernandez, Ho, Hogg,
  Holtzman, Honscheid, Inada, Ivans, Jiang, Jiang, Johnson, Jordan, Jordan,
  Kauffmann, Kazin, Kirkby, Klaene, Kneib, Knapp, Kochanek, Koesterke,
  Kollmeier, Kron, Lang, Lawler, Goff, Lee, Lee, Leisenring, Lin, Liu, Long,
  Loomis, Lucatello, Lundgren, Lupton, Ma, Ma, MacDonald, Mack, Mahadevan,
  Maia, Malanushenko, Malanushenko, Majewski, Makler, Mandelbaum, Maraston,
  Margala, Maseman, Masters, McBride, McDonald, McGreer, McMahon, Requejo,
  Menard, Miralda-Escude, Morrison, Mullally, Muna, Murayama, Myers, Naugle,
  Neto, Nguyen, Nichol, Nidever, O'Connell, Ogando, Olmstead, Oravetz,
  Padmanabhan, Paegert, Palanque-Delabrouille, Pan, Pandey, Parejko, Paris,
  Pellegrini, Pepper, Percival, Petitjean, Pfaffenberger, Pforr, Phleps,
  Pichon, Pieri, Prada, Price-Whelan, Raddick, Ramos, Ryle, Reid, Rich,
  Richards, Rieke, Rieke, Rix, Robin, Rocha-Pinto, Rockosi, Roe, Rollinde,
  Ross, Ross, Rossetto, Sanchez, Santiago, Sayres, Schiavon, Schlegel,
  Schlesinger, Schmidt, Schneider, Sellgren, Shelden, Sheldon, Shetrone, Shu,
  Silverman, Simmerer, Simmons, Sivarani, Skrutskie, Slosar, Smee, Smith,
  Snedden, Stassun, Steele, Steinmetz, Stockett, Stollberg, Strauss, Tanaka,
  Thakar, Thomas, Tinker, Tofflemire, Tojeiro, Tremonti, Magana, Verde, Vogt,
  Wake, Wan, Wang, Weaver, White, White, Wilson, Wisniewski, Wood-Vasey, Yanny,
  Yasuda, Yeche, York, Young, Zasowski, Zehavi, \& Zhao}]{Eisenstein:2011}
Eisenstein, D.~J., {et~al.} 2011, ArXiv e-prints

\bibitem[{{Falk} {et~al.}(1993){Falk}, {Rangarajan}, \&
  {Srednicki}}]{Falk:1993}
{Falk}, T., {Rangarajan}, R., \& {Srednicki}, M. 1993, \apjl, 403, L1

\bibitem[{{Fosalba} \& {Gaztanaga}(1998)}]{Fosalba:1998}
{Fosalba}, P., \& {Gaztanaga}, E. 1998, \mnras, 301, 503

\bibitem[{{Frieman} \& {Gaztanaga}(1994)}]{Frieman:1994}
{Frieman}, J.~A., \& {Gaztanaga}, E. 1994, \apj, 425, 392

\bibitem[{{Fry}(1985)}]{Fry:1985}
{Fry}, J.~N. 1985, \apj, 289, 10

\bibitem[{{Fry} \& {Gaztanaga}(1993)}]{Fry:1993}
{Fry}, J.~N., \& {Gaztanaga}, E. 1993, \apj, 413, 447

\bibitem[{{Fry} \& {Gaztanaga}(1994)}]{Fry:1994a}
---. 1994, \apj, 425, 1

\bibitem[{{Fry} \& {Scherrer}(1994)}]{Fry:1994}
{Fry}, J.~N., \& {Scherrer}, R.~J. 1994, \apj, 429, 36

\bibitem[{{Gangui} {et~al.}(1994){Gangui}, {Lucchin}, {Matarrese}, \&
  {Mollerach}}]{Gangui:1994}
{Gangui}, A., {Lucchin}, F., {Matarrese}, S., \& {Mollerach}, S. 1994, \apj,
  430, 447

\bibitem[{{Gaztanaga}(1992)}]{Gaztanaga:1992}
{Gaztanaga}, E. 1992, \apjl, 398, L17

\bibitem[{{Ghigna} {et~al.}(1996){Ghigna}, {Bonometto}, {Guzzo}, {Giovanelli},
  {Haynes}, {Klypin}, \& {Primack}}]{Ghigna:1996}
{Ghigna}, S., {Bonometto}, S.~A., {Guzzo}, L., {Giovanelli}, R., {Haynes},
  M.~P., {Klypin}, A., \& {Primack}, J.~R. 1996, \apj, 463, 395

\bibitem[{{Giannantonio} \& {Porciani}(2010)}]{Giannantonio:2010}
{Giannantonio}, T., \& {Porciani}, C. 2010, \prd, 81, 063530

\bibitem[{{Giannantonio} {et~al.}(2012){Giannantonio}, {Porciani}, {Carron},
  {Amara}, \& {Pillepich}}]{Giannantonio:2012}
{Giannantonio}, T., {Porciani}, C., {Carron}, J., {Amara}, A., \& {Pillepich},
  A. 2012, \mnras, 422, 2854

\bibitem[{{Giannantonio} {et~al.}(2013){Giannantonio}, {Ross}, {Percival},
  {Crittenden}, {Bacher}, {Kilbinger}, {Nichol}, \&
  {Weller}}]{Giannantonio:2013}
{Giannantonio}, T., {Ross}, A.~J., {Percival}, W.~J., {Crittenden}, R.,
  {Bacher}, D., {Kilbinger}, M., {Nichol}, R., \& {Weller}, J. 2013, ArXiv
  e-prints

\bibitem[{{Guo} {et~al.}(2013){Guo}, {Zehavi}, {Zheng}, {Weinberg}, {Berlind},
  {Blanton}, {Chen}, {Eisenstein}, {Ho}, {Kazin}, {Manera}, {Maraston},
  {McBride}, {Nuza}, {Padmanabhan}, {Parejko}, {Percival}, {Ross}, {Ross},
  {Samushia}, {S{\'a}nchez}, {Schlegel}, {Schneider}, {Skibba}, {Swanson},
  {Tinker}, {Tojeiro}, {Wake}, {White}, {Bahcall}, {Bizyaev}, {Brewington},
  {Bundy}, {da Costa}, {Ebelke}, {Malanushenko}, {Malanushenko}, {Oravetz},
  {Rossi}, {Simmons}, {Snedden}, {Streblyanska}, \& {Thomas}}]{Guo:2013}
{Guo}, H., {et~al.} 2013, \apj, 767, 122

\bibitem[{{Guth}(1981)}]{Guth:1981}
{Guth}, A.~H. 1981, \prd, 23, 347

\bibitem[{{Hoyle} {et~al.}(2000){Hoyle}, {Szapudi}, \& {Baugh}}]{Hoyle:2000}
{Hoyle}, F., {Szapudi}, I., \& {Baugh}, C.~M. 2000, \mnras, 317, L51

\bibitem[{{Hui} \& {Gazta{\~n}aga}(1999)}]{Hui:1999}
{Hui}, L., \& {Gazta{\~n}aga}, E. 1999, \apj, 519, 622

\bibitem[{{Juszkiewicz} \& {Bouchet}(1992)}]{Juszkiewicz:1992}
{Juszkiewicz}, R., \& {Bouchet}, F.~R. 1992, in Distribution of Matter in the
  Universe, ed. G.~A.~M. .~D. Gerbal, 301--310

\bibitem[{{Juszkiewicz} {et~al.}(1993){Juszkiewicz}, {Bouchet}, \&
  {Colombi}}]{Juszkiewicz:1993}
{Juszkiewicz}, R., {Bouchet}, F.~R., \& {Colombi}, S. 1993, \apjl, 412, L9

\bibitem[{{Kazin} {et~al.}(2010){Kazin}, {Blanton}, {Scoccimarro}, {McBride},
  {Berlind}, {Bahcall}, {Brinkmann}, {Czarapata}, {Frieman}, {Kent},
  {Schneider}, \& {Szalay}}]{Kazin:2010}
{Kazin}, E.~A., {et~al.} 2010, \apj, 710, 1444

\bibitem[{{Kim} \& {Strauss}(1998)}]{Kim:1998}
{Kim}, R.~S., \& {Strauss}, M.~A. 1998, \apj, 493, 39

\bibitem[{{Komatsu} \& {Spergel}(2001)}]{Komatsu:2001}
{Komatsu}, E., \& {Spergel}, D.~N. 2001, \prd, 63, 063002

\bibitem[{{Lahav} {et~al.}(1993){Lahav}, {Itoh}, {Inagaki}, \&
  {Suto}}]{Lahav:1993}
{Lahav}, O., {Itoh}, M., {Inagaki}, S., \& {Suto}, Y. 1993, \apj, 402, 387

\bibitem[{{Lam} \& {Sheth}(2009)}]{Lam:2009a}
{Lam}, T.~Y., \& {Sheth}, R.~K. 2009, \mnras, 395, 1743

\bibitem[{{Lam} {et~al.}(2009){Lam}, {Sheth}, \& {Desjacques}}]{Lam:2009b}
{Lam}, T.~Y., {Sheth}, R.~K., \& {Desjacques}, V. 2009, \mnras, 399, 1482

\bibitem[{{Lucchin} {et~al.}(1994){Lucchin}, {Matarrese}, {Melott}, \&
  {Moscardini}}]{Lucchin:1994}
{Lucchin}, F., {Matarrese}, S., {Melott}, A.~L., \& {Moscardini}, L. 1994,
  \apj, 422, 430

\bibitem[{{Luo} \& {Schramm}(1993)}]{Luo:1993}
{Luo}, X., \& {Schramm}, D.~N. 1993, \apj, 408, 33

\bibitem[{{Maldacena}(2003)}]{Maldacena:2003}
{Maldacena}, J. 2003, Journal of High Energy Physics, 5, 13

\bibitem[{{Manera} \& {Gazta{\~n}aga}(2011)}]{Manera:2011}
{Manera}, M., \& {Gazta{\~n}aga}, E. 2011, \mnras, 415, 383

\bibitem[{{Mar{\'{\i}}n}(2011)}]{Marin:2011}
{Mar{\'{\i}}n}, F. 2011, \apj, 737, 97

\bibitem[{{Marinoni} {et~al.}(2005){Marinoni}, {Le F{\`e}vre}, {Meneux},
  {Iovino}, {Pollo}, {Ilbert}, {Zamorani}, {Guzzo}, {Mazure}, {Scaramella},
  {Cappi}, {McCracken}, {Bottini}, {Garilli}, {Le Brun}, {Maccagni}, {Picat},
  {Scodeggio}, {Tresse}, {Vettolani}, {Zanichelli}, {Adami}, {Arnouts},
  {Bardelli}, {Blaizot}, {Bolzonella}, {Charlot}, {Ciliegi}, {Contini},
  {Foucaud}, {Franzetti}, {Gavignaud}, {Marano}, {Mathez}, {Merighi},
  {Paltani}, {Pell{\`o}}, {Pozzetti}, {Radovich}, {Zucca}, {Bondi},
  {Bongiorno}, {Busarello}, {Colombi}, {Cucciati}, {Lamareille}, {Mellier},
  {Merluzzi}, {Ripepi}, \& {Rizzo}}]{Marinoni:2005}
{Marinoni}, C., {et~al.} 2005, \aap, 442, 801

\bibitem[{{McBride} {et~al.}(2009){McBride}, {Berlind}, {Scoccimarro},
  {Wechsler}, {Busha}, {Gardner}, \& {van den Bosch}}]{McBride:2009}
{McBride}, C., {Berlind}, A., {Scoccimarro}, R., {Wechsler}, R., {Busha}, M.,
  {Gardner}, J., \& {van den Bosch}, F. 2009, in Bulletin of the American
  Astronomical Society, Vol.~41, American Astronomical Society Meeting
  Abstracts \#213, 425.06

\bibitem[{{Nuza} {et~al.}(2013){Nuza}, {S{\'a}nchez}, {Prada}, {Klypin},
  {Schlegel}, {Gottl{\"o}ber}, {Montero-Dorta}, {Manera}, {McBride}, {Ross},
  {Angulo}, {Blanton}, {Bolton}, {Favole}, {Samushia}, {Montesano}, {Percival},
  {Padmanabhan}, {Steinmetz}, {Tinker}, {Skibba}, {Schneider}, {Guo}, {Zehavi},
  {Zheng}, {Bizyaev}, {Malanushenko}, {Malanushenko}, {Oravetz}, {Oravetz}, \&
  {Shelden}}]{Nuza:2013}
{Nuza}, S.~E., {et~al.} 2013, \mnras, 432, 743

\bibitem[{{P{\'a}pai} \& {Szapudi}(2010)}]{P'apai:2010}
{P{\'a}pai}, P., \& {Szapudi}, I. 2010, \apj, 725, 2078

\bibitem[{{Parejko} {et~al.}(2013){Parejko}, {Sunayama}, {Padmanabhan}, {Wake},
  {Berlind}, {Bizyaev}, {Blanton}, {Bolton}, {van den Bosch}, {Brinkmann},
  {Brownstein}, {da Costa}, {Eisenstein}, {Guo}, {Kazin}, {Maia},
  {Malanushenko}, {Maraston}, {McBride}, {Nichol}, {Oravetz}, {Pan},
  {Percival}, {Prada}, {Ross}, {Ross}, {Schlegel}, {Schneider}, {Simmons},
  {Skibba}, {Tinker}, {Tojeiro}, {Weaver}, {Wetzel}, {White}, {Weinberg},
  {Thomas}, {Zehavi}, \& {Zheng}}]{Parejko:2013}
{Parejko}, J.~K., {et~al.} 2013, \mnras, 429, 98

\bibitem[{{Peebles}(1980)}]{Peebles:1980}
{Peebles}, P.~J.~E. 1980, The large-scale structure of the universe (Princeton
  University Press, Princeton, New Jersey)

\bibitem[{{Planck Collaboration} {et~al.}(2013){Planck Collaboration}, {Ade},
  {Aghanim}, {Armitage-Caplan}, {Arnaud}, {Ashdown}, {Atrio-Barandela},
  {Aumont}, {Baccigalupi}, {Banday}, \& et~al.}]{PlanckCollaboration:2013}
{Planck Collaboration} {et~al.} 2013, arXiv1303.5084

\bibitem[{{Ross} {et~al.}(2008){Ross}, {Brunner}, \& {Myers}}]{Ross:2008}
{Ross}, A.~J., {Brunner}, R.~J., \& {Myers}, A.~D. 2008, \apj, 682, 737

\bibitem[{{Ross} {et~al.}(2013){Ross}, {Percival}, {Carnero}, {Zhao}, {Manera},
  {Raccanelli}, {Aubourg}, {Bizyaev}, {Brewington}, {Brinkmann}, {Brownstein},
  {Cuesta}, {da Costa}, {Eisenstein}, {Ebelke}, {Guo}, {Hamilton},
  {Maga{\~n}a}, {Malanushenko}, {Malanushenko}, {Maraston}, {Montesano},
  {Nichol}, {Oravetz}, {Pan}, {Prada}, {S{\'a}nchez}, {Samushia}, {Schlegel},
  {Schneider}, {Seo}, {Sheldon}, {Simmons}, {Snedden}, {Swanson}, {Thomas},
  {Tinker}, {Tojeiro}, \& {Zehavi}}]{Ross:2013}
{Ross}, A.~J., {et~al.} 2013, \mnras, 428, 1116

\bibitem[{{Salopek} \& {Bond}(1990)}]{Salopek:1990}
{Salopek}, D.~S., \& {Bond}, J.~R. 1990, \prd, 42, 3936

\bibitem[{{Saunders} {et~al.}(1991){Saunders}, {Frenk}, {Rowan-Robinson},
  {Lawrence}, \& {Efstathiou}}]{Saunders:1991}
{Saunders}, W., {Frenk}, C., {Rowan-Robinson}, M., {Lawrence}, A., \&
  {Efstathiou}, G. 1991, \nat, 349, 32

\bibitem[{{Scherrer} \& {Schaefer}(1995)}]{Scherrer:1995}
{Scherrer}, R.~J., \& {Schaefer}, R.~K. 1995, \apj, 446, 44

\bibitem[{{Scoccimarro}(1998)}]{Roman:1998}
{Scoccimarro}, R. 1998, \mnras, 299, 1097

\bibitem[{{Scoccimarro} \& {Frieman}(1996)}]{Scoccimarro:1996}
{Scoccimarro}, R., \& {Frieman}, J. 1996, \apjs, 105, 37

\bibitem[{{Scoccimarro} {et~al.}(2012){Scoccimarro}, {Hui}, {Manera}, \&
  {Chan}}]{Scoccimarro:2012}
{Scoccimarro}, R., {Hui}, L., {Manera}, M., \& {Chan}, K.~C. 2012, \prd, 85,
  083002

\bibitem[{Scoccimarro {et~al.}(2004)Scoccimarro, Sefusatti, \&
  Zaldarriaga}]{Scoccimarro:2004}
Scoccimarro, R., Sefusatti, E., \& Zaldarriaga, M. 2004, Phys.Rev. D, 69,
  103513

\bibitem[{{Sefusatti}(2009)}]{Sefusatti:2009}
{Sefusatti}, E. 2009, \prd, 80, 123002

\bibitem[{{Sefusatti} {et~al.}(2006){Sefusatti}, {Crocce}, {Pueblas}, \&
  {Scoccimarro}}]{Sefusatti:2006}
{Sefusatti}, E., {Crocce}, M., {Pueblas}, S., \& {Scoccimarro}, R. 2006, \prd,
  74, 023522

\bibitem[{{Sefusatti} \& {Komatsu}(2007)}]{Sefusatti:2007}
{Sefusatti}, E., \& {Komatsu}, E. 2007, \prd, 76, 083004

\bibitem[{{Senatore} {et~al.}(2010){Senatore}, {Smith}, \&
  {Zaldarriaga}}]{Senatore:2010}
{Senatore}, L., {Smith}, K.~M., \& {Zaldarriaga}, M. 2010, J. Cosmol.
  Astropart. Phys., 1, 28, original template of orthogonal shape
  non-Gaussianity

\bibitem[{{Slosar} {et~al.}(2008){Slosar}, {Hirata}, {Seljak}, {Ho}, \&
  {Padmanabhan}}]{Slosar:2008}
{Slosar}, A., {Hirata}, C., {Seljak}, U., {Ho}, S., \& {Padmanabhan}, N. 2008,
  J. Cosmol. Astropart. Phys., 8, 31

\bibitem[{{Springel}(2005)}]{Springel:2005}
{Springel}, V. 2005, \mnras, 364, 1105

\bibitem[{{Szapudi} {et~al.}(2000){Szapudi}, {Branchini}, {Frenk}, {Maddox}, \&
  {Saunders}}]{Szapudi:2000p}
{Szapudi}, I., {Branchini}, E., {Frenk}, C.~S., {Maddox}, S., \& {Saunders}, W.
  2000, \mnras, 318, L45

\bibitem[{{Szapudi} {et~al.}(2002){Szapudi}, {Frieman}, {Scoccimarro},
  {Szalay}, {Connolly}, {Dodelson}, {Eisenstein}, {Gunn}, {Johnston}, {Kent},
  {Loveday}, {Meiksin}, {Nichol}, {Scranton}, {Stebbins}, {Vogeley}, {Annis},
  {Bahcall}, {Brinkman}, {Csabai}, {Doi}, {Fukugita}, {Ivezi{\'c}}, {Kim},
  {Knapp}, {Lamb}, {Lee}, {Lupton}, {McKay}, {Munn}, {Peoples}, {Pier},
  {Rockosi}, {Schlegel}, {Stoughton}, {Tucker}, {Yanny}, \&
  {York}}]{Szapudi:2002}
{Szapudi}, I., {et~al.} 2002, \apj, 570, 75

\bibitem[{{Verde} {et~al.}(2001){Verde}, {Jimenez}, {Kamionkowski}, \&
  {Matarrese}}]{Verde:2001}
{Verde}, L., {Jimenez}, R., {Kamionkowski}, M., \& {Matarrese}, S. 2001,
  \mnras, 325, 412

\bibitem[{{Verde} {et~al.}(2000){Verde}, {Wang}, {Heavens}, \&
  {Kamionkowski}}]{Verde:2000}
{Verde}, L., {Wang}, L., {Heavens}, A.~F., \& {Kamionkowski}, M. 2000, \mnras,
  313, 141

\bibitem[{{Weinberg} \& {Cole}(1992)}]{Weinberg:1992}
{Weinberg}, D.~H., \& {Cole}, S. 1992, \mnras, 259, 652

\bibitem[{{York} {et~al.}(2000){York}, {Adelman}, {Anderson}, {Anderson},
  {Annis}, {Bahcall}, {Bakken}, {Barkhouser}, {Bastian}, {Berman}, {Boroski},
  {Bracker}, {Briegel}, {Briggs}, {Brinkmann}, {Brunner}, {Burles}, {Carey},
  {Carr}, {Castander}, {Chen}, {Colestock}, {Connolly}, {Crocker}, {Csabai},
  {Czarapata}, {Davis}, {Doi}, {Dombeck}, {Eisenstein}, {Ellman}, {Elms},
  {Evans}, {Fan}, {Federwitz}, {Fiscelli}, {Friedman}, {Frieman}, {Fukugita},
  {Gillespie}, {Gunn}, {Gurbani}, {de Haas}, {Haldeman}, {Harris}, {Hayes},
  {Heckman}, {Hennessy}, {Hindsley}, {Holm}, {Holmgren}, {Huang}, {Hull},
  {Husby}, {Ichikawa}, {Ichikawa}, {Ivezi{\'c}}, {Kent}, {Kim}, {Kinney},
  {Klaene}, {Kleinman}, {Kleinman}, {Knapp}, {Korienek}, {Kron}, {Kunszt},
  {Lamb}, {Lee}, {Leger}, {Limmongkol}, {Lindenmeyer}, {Long}, {Loomis},
  {Loveday}, {Lucinio}, {Lupton}, {MacKinnon}, {Mannery}, {Mantsch}, {Margon},
  {McGehee}, {McKay}, {Meiksin}, {Merelli}, {Monet}, {Munn}, {Narayanan},
  {Nash}, {Neilsen}, {Neswold}, {Newberg}, {Nichol}, {Nicinski}, {Nonino},
  {Okada}, {Okamura}, {Ostriker}, {Owen}, {Pauls}, {Peoples}, {Peterson},
  {Petravick}, {Pier}, {Pope}, {Pordes}, {Prosapio}, {Rechenmacher}, {Quinn},
  {Richards}, {Richmond}, {Rivetta}, {Rockosi}, {Ruthmansdorfer}, {Sandford},
  {Schlegel}, {Schneider}, {Sekiguchi}, {Sergey}, {Shimasaku}, {Siegmund},
  {Smee}, {Smith}, {Snedden}, {Stone}, {Stoughton}, {Strauss}, {Stubbs},
  {SubbaRao}, {Szalay}, {Szapudi}, {Szokoly}, {Thakar}, {Tremonti}, {Tucker},
  {Uomoto}, {Vanden Berk}, {Vogeley}, {Waddell}, {Wang}, {Watanabe},
  {Weinberg}, {Yanny}, \& {Yasuda}}]{York:2000}
{York}, D.~G., {et~al.} 2000, \aj, 120, 1579

\bibitem[{{Zehavi} {et~al.}(2005){Zehavi}, {Eisenstein}, {Nichol}, {Blanton},
  {Hogg}, {Brinkmann}, {Loveday}, {Meiksin}, {Schneider}, \&
  {Tegmark}}]{Zehavi:2005}
{Zehavi}, I., {et~al.} 2005, \apj, 621, 22

\end{thebibliography}

\end{document}